\documentclass[epj]{svjour}
\usepackage{graphics}
\usepackage{hyperref}
\usepackage{amsmath}
\usepackage{amssymb}
\usepackage{bm}
\usepackage{braket}
\usepackage{mathrsfs}
\usepackage[compress]{cite}
\begin{document}
\title{Exploratory study on the masses of odd-$Z$ nuclei and $r$-process simulation based on the deformed relativistic Hartree-Bogoliubov theory in continuum }
\author{Cong Pan\inst{1} \and Yaochen Yang \inst{2} \and Xiaofei Jiang \inst{2} \and Xin-Hui Wu \inst{3}
\thanks{e-mail: wuxinhui@fzu.edu.cn (corresponding author)}%
}                     
\institute{Department of Physics, Anhui Normal University, Wuhu 241000, China 
\and State Key Laboratory of Nuclear Physics and Technology, School of Physics, Peking University, Beijing 100871, China 
\and Department of Physics, Fuzhou University, Fuzhou 350108, Fujian, China }
\date{Received: date / Revised version: date}
%
\abstract{
Nuclear masses of exotic nuclei are important for both nuclear physics and astrophysics.
The deformed relativistic Hartree-Bogoliubov theory in continuum (DRHBc) is capable of providing proper descriptions for exotic nuclei by simultaneously including deformation, pairing correlation and continuum effects, and a mass table of even-$Z$ nuclei with $8 \leqslant Z \leqslant 120$ has been developed based on the DRHBc theory.
This work employs a methodology to estimate the masses of odd nuclei using neighboring even nuclei's masses and microscopic pairing gaps, and the performance of microscopic pairing gaps are validated by comparing with empirical ones.
Combining the DRHBc masses of even-$Z$ nuclei and the estimated masses of odd-$Z$ nuclei, a pseudo DRHBc mass table is developed, with the root-mean-square (rms) deviation from available mass data $\sigma=1.47$ MeV. 
Then this mass table is employed in the $r$-process simulation; results show that the differences in the details of pairing gaps do not yield qualitative discrepancy in $r$-process abundances, while the deformation effects can influence the $r$-process path and thus affect the $r$-process abundance. 
In particular, the nuclear shape transitions can even lead to the discontinuity of the $r$-process path, suggesting that incorporating triaxiality or beyond-mean-field effects would be valuable for further improvement.
%
} 
\maketitle

\section{Introduction} \label{intro}

Nuclear masses are important for both nuclear physics \cite{Yamaguchi2021PPNP} and astrophysics~\cite{Kajino2019PPNP}.
They reflect a variety of underlying physical effects of nuclear quantum many-body systems, and can be used to extract nuclear structure information, e.g., nuclear deformation \cite{Roubin2017PRC}, shell effects \cite{Ramirez2012Science}, and nuclear force \cite{Wienholtz2013Nature}.
They also determine the reaction energies for all nuclear reactions, which are important in understanding the energy production in stars \cite{Bethe1939PR} and the study of nucleosynthesis \cite{Mumpower2015PRC}.
However, experimentally, the masses of about 2500 nuclei have been measured so far \cite{Wang2021CPC}.

The rapid neutron capture process ($r$-process) has been believed to be responsible for the nucleosynthesis of about half of the elements heavier than iron for more than half a century \cite{Burbidge1957RMP,Thielemann2017ARNPS,Cowan2021RMP}. 
The understanding of $r$-process is still affected by the uncertainties in our knowledge of both nuclear physics quantities and astrophysical conditions~\cite{Zhao2019APJ,Kajino2019PPNP,Cowan2021RMP,Arcones2023AAR,Wu2023Sci.Bull.}.
The masses of neutron-rich nuclei are crucial to the $r$-process studies, as they are needed to extract the reaction energies that go into the calculations of all involved nuclear reaction rates, i.e., neutron separation energies ($S_n$) for neutron-capture cross sections and $\beta$-decay $Q$-values ($Q_{\beta}$) for $\beta$-decay half-lives ($T_{1/2}$)~\cite{Mumpower2016PPNP,Wu2022APJ, Huang2025ApJ}.
Even in the classical $r$-process model with waiting-point approximation (WPA), the use of different nuclear mass models would affect the equilibrium between neutron capture and photodisintegration reactions to predict $r$-process path and also the $Q$-values of $\beta$-decay half-lives, and thus affect the final calculated abundance~\cite{Jiang2021APJ}.
Nuclear masses of most neutron-rich nuclei, especially the exotic nuclei near drip lines, remain beyond the current experimental capabilities even in the foreseeable future, due to the difficulties in production, separation, and detection. 
Therefore, theoretical models with reliable predictive power are essential. 

Many efforts have been made to theoretically describe nuclear masses, including macroscopic-microscopic models \cite{Pearson1996PLB,Wang2014PLB,Moller2016ADNDT,Koura2005PTP}, 
microscopic models \cite{Goriely2009PRL_Skyrme,Goriely2009PRL_Gogny,Pena-Arteaga2016EPJA,Xia2018ADNDT,Yang2021PRC,Zhang2022ADNDT},
and machine-learning approaches \cite{Utama2016PRC,Neufcourt2018PRC,Neufcourt2019PRL,Wu2020PRC,Wu2021PLB,Niu2022PRC,Du2023CPC,Wu2024PRC,Wu2024PRC_AKRR,Wu2024SC,Guo2024PRC}.
Microscopic models are usually believed to have a better reliability of extrapolation \cite{Zhao2012PRCmass,Zhang2021PRC,He2024PRC}. 
This is particularly necessary for the mass predictions for the exotic nuclei, which are far away from the experimentally known region and play important roles in the $r$-process path.

The exotic nuclei involved in the $r$-process path are extremely weakly bound, and their Fermi energies are very close to the continuum threshold. 
For these nuclei, the pairing interaction can scatter nucleons from bound states to the resonant ones in the continuum, and the density could become more diffuse due to this coupling to the continuum \cite{Meng2006PPNP}. 
The stability of exotic nuclei and even the position of the drip line might be influenced, which is the so-called continuum effect. 
Therefore, the effects of pairing correlation and the coupling to continuum should be considered properly \cite{Dobaczewski1984NPA,Dobaczewski1996PRC,Grasso2001PRC,Michel2008PRC,Pei2011PRC,Zhang2013PRC} in the description for exotic nuclei near drip lines.  
This is important for the studies of $r$-process in extremely high-neutron-density environments, e.g., the neutron star mergers. 
The relativistic continuum Hartree-Bogoliubov (RCHB) theory \cite{Meng1996PRL,Meng1998NPA}, by assuming spherical symmetry and solving the relativistic Hartree-Bogoliubov (RHB) equations for nucleons in the coordinate space, takes into account pairing correlations and continuum effect in a microscopic and self-consistent way, and has achieved great success in describing both stable and exotic nuclei \cite{Meng1998PRL,Meng2006PPNP}.
	
Based on the RCHB theory, the first nuclear mass table including continuum effect was constructed \cite{Xia2018ADNDT}, and the continuum effect on the limits of the nuclear landscape was studied. 
However, the accuracy of the RCHB mass table is limited due to the assumed spherical symmetry.

Most nuclei in the nuclear chart, except for those with shell closures, exhibit deformed shapes.
In order to properly describe deformed exotic nuclei, the deformed relativistic Hartree-Bogoliubov theory in continuum (DRHBc) was developed by solving the deformed RHB equations in a Dirac Woods-Saxon basis, and can treat the effects of deformation, pairing correlations, and continuum simultaneously \cite{Zhou2010PRC,Li2012PRC}.
As the advantages of the RCHB theory are inherited and the deformation degree of freedom is further included, the DRHBc theory has been successfully applied in a variety of studies on exotic nuclei, including the halo structures \cite{Sun2018PLB,Yang2021PRL,Sun2021PRC,Zhong2022SCP,Zhang2023PRC_Na,Zhang2023PLB,Pan2024PLB,Zhang2024PRC_Al}, peninsulas beyond the neutron drip line \cite{Zhang2021PRC,Pan2021PRC,He2021CPC,He2024PRC}, evolution of nuclear shape and size \cite{Choi2022PRC,Kim2022PRC,Mun2023PLB,Guo2023PRC,Pan2025PRC}, half-life estimation for proton emission and $\alpha$-decay \cite{Xiao2023PLB,Choi2024PRC,Lu2024PLB}, etc. 
The predictive power of the DRHBc theory for nuclear mass has been examined in Refs.~\cite{Zhang2021PRC,He2024PRC} by taking the even-even superheavy nuclei with $102 \leqslant Z \leqslant 120$ as examples. 
Recently, based on the DRHBc theory with the relativistic density functional PC-PK1 and the density-dependent zero-range pairing force, a nuclear mass table for the nuclei with $8 \leqslant Z \leqslant 120$ is in progress \cite{Zhang2020PRC,Pan2022PRC}.
The DRHBc mass table is expected to provide a microscopic mass input with the effects of deformation and continuum simultaneously included. 
Up to now, the even-$Z$ part of the DRHBc mass table has been developed \cite{Zhang2022ADNDT,Guo2024ADNDT}. 

Since the results for odd-$Z$ nuclei are unavailable in the current DRHBc mass table, our work aims to systematically estimate their masses based on the available results of even-$Z$ nuclei, and then utilize them to perform $r$-process simulations. 
This article is structured as follows: 
The theoretical frameworks for the DRHBc theory and classical $r$-process are briefly introduced in Section \ref{Stheory}, 
the examination of the methodology and the estimation of the masses of odd-$Z$ nuclei are presented in Section \ref{Smass}, the $r$-process simulation based on the obtained masses and the impact of the deformation effects are discussed in Section \ref{Srpro}, and finally, a summary is given in Section \ref{Ssum}. 

\section{ Theoretical framework } \label{Stheory}

\subsection{ The DRHB{c} theory }

The details of the DRHBc theory have been illustrated in Refs.~\cite{Li2012PRC,Zhang2020PRC,Pan2022PRC}. 
In this Section, we just present a brief theoretical framework. 
In the DRHBc theory, the motion of nucleons are microscopically described by the relativistic Hartree-Bogoliubov (RHB) equation \cite{Kucharek1991ZPA}, 
\begin{equation} 
	\label{RHB}
	\begin{pmatrix} \hat{h}_D - \lambda_\tau & \hat{\Delta} \\ -\hat{\Delta}^* & -\hat{h}_D^* + \lambda_\tau \end{pmatrix} 
	\begin{pmatrix} U_k \\ V_k \end{pmatrix} = E_k \begin{pmatrix} U_k \\ V_k \end{pmatrix}, 
\end{equation}
where $\hat{h}_D$ is the Dirac Hamiltonian, $\hat{\Delta}$ is the pairing potential, $E_k$ is the quasiparticle energy, $U_k$ and $V_k$ are the quasiparticle wave functions, and $\lambda_\tau$ is the Fermi energy of neutron or proton $(\tau = n,p)$. 

In the coordinate space, the Dirac Hamiltonian $\hat{h}_D$ reads 
\begin{equation}
	h_D(\bm{r}) = \bm{\alpha}\cdot\bm{p} + V(\bm{r}) + \beta[M+S(\bm{r})],
\end{equation}
where $S(\bm{r})$ and $V(\bm{r})$ are scalar and vector potentials, respectively.
The pairing potential $\hat{\Delta}$ reads
\begin{equation}
	\Delta(\bm{r}_1,\bm{r}_2) = V^{pp}(\bm{r}_1,\bm{r}_2) \kappa(\bm{r}_1,\bm{r}_2),
\end{equation}
where $V^{pp}$ is the pairing force, and $\kappa$ is the pairing tensor \cite{Ring1980NMBP}. 
In this work, the density-dependent zero-range pairing force
\begin{equation}
	\label{eVpp}
	V^{pp}(\bm{r}_1,\bm{r}_2) = V_0 \frac{1}{2} (1-P^\sigma) \delta(\bm{r}_1 - \bm{r}_2) \left(1 - \frac{\rho(\bm{r}_1)}{\rho_{\text{sat}}}\right) 
\end{equation}
is adopted. 

In the DRHBc theory, the axial deformation and spatial reflection symmetry are assumed, and the potentials and densities can be expanded in terms of the Legendre polynomials,
\begin{equation}
	\label{elam}
	f(\bm{r}) = \sum_\lambda f_\lambda(r) P_\lambda(\cos\theta), \quad \lambda = 0,2,4,\dots, \lambda_{\max}.
\end{equation}
For the exotic nuclei close to drip lines, the continuum effect should be taken into account properly \cite{Dobaczewski1996PRC,Meng2006PPNP}. 
For this purpose, the deformed RHB equations \eqref{RHB} are solved in a spherical Dirac Woods-Saxon basis \cite{Zhou2003PRC,Zhang2022PRC}, which can properly describe the asymptotic behavior of the density distribution at a large $r$ for exotic nuclei. 
Since the odd nucleon breaks the time-reversal symmetry and leads to nonzero nucleon currents \cite{Pan2024PLB}, to avoid such complexity, the time-reversal symmetry is assumed with the equal filling approximation \cite{Perez-Martin2008PRC,Li2012CPL}. 

For a nucleus with odd number of neutron or proton, the blocking effect of the unpaired nucleon(s) needs to be considered \cite{Ring1980NMBP}. 
Practically, this can be realized by the exchange of quasiparticle wavefunctions $(U_{k_b}, V_{k_b}) \leftrightarrow (V_{k_b}^*, U_{k_b}^*)$ and that of the energy $E_{k_b} \leftrightarrow -E_{k_b}$ for Eq.~\eqref{RHB}, where $k_b$ refers to the blocked orbital for the odd nucleon \cite{Li2012CPL,Pan2022PRC}. 

After self-consistently solving the RHB equations, the expectation values such as binding energy, quadrupole deformation, root-mean-square radii, etc., can be calculated \cite{Li2012PRC,Zhang2020PRC,Pan2022PRC}. 
The canonical basis $\ket{\psi_i}$ is obtained by the following diagonalization \cite{Ring1980NMBP}:
\begin{equation}
    \hat{\rho} \ket{\psi_i} = v_i^2 \ket{\psi_i}, 
\end{equation}
where $\hat{\rho}$ is the density matrix, and $v_i^2$ is the corresponding occupation probability of $\ket{\psi_i}$. 
The single-particle energy in the canonical basis is obtained as $\epsilon_i = \braket{\psi_i|\hat{h}_D|\psi_i}$. 
The pairing gap $\Delta_i$ is calculated by 
\begin{equation}
    \Delta_i = \frac{2u_i v_i}{u_i^2 - v_i^2}(\epsilon_i - \lambda_\tau), 
\end{equation}
where the parameter $u_i$ is obtained from $u_i^2 + v_i^2 = 1$. 
The average pairing gap defined by \cite{Dobaczewski1996PRC,Agbemava2014PRC}: 
\begin{equation}
    \label{eDel}
    \Delta = \frac{\sum_i v_i^2 \Delta_i}{\sum_i v_i^2} ,
\end{equation}
is an order parameter describing the phase transition from a normal fluid to a superfluid \cite{Afanasjev2015PRC}. 
It should also be mentioned that for deformed nuclei, due to the breaking of the rotational symmetry, one should also consider the rotational correction energy \cite{Zhao2010PRC,Zhang2020PRC}, 
\begin{equation} \label{Erot}
	E_{\mathrm{rot}} = -\frac{ \braket{\hat{\bm{J}}^2} }{2 \mathscr{I} } ,
\end{equation}
where $\hat{\bm{J}}$ is the total angular momentum operator, and $\mathscr{I}$ is the moment of inertia calculated by the Inglis-Belyaev formula with the cranking approximation \cite{Ring1980NMBP}. 
In Refs. \cite{Zhang2022ADNDT,Guo2024ADNDT}, it was mentioned that the cranking approximation is not suitable for (near-)spherical nuclei, and therefore, the $E_{\text{rot}}$ values of the nuclei with $|\beta_2|<0.05$ are taken as zero in the DRHBc mass table. 
The two-dimensional collective Hamiltonian method \cite{Sun2022CPC} could better consider the correction energies for the near-spherical nuclei \cite{Zhang2023PRC_2DCH} and is expected to provide an alternative treatment for future improvement for the DRHBc mass table.

For completeness, here we briefly list the numerical conditions adopted in the DRHBc mass table calculations \cite{Zhang2020PRC,Pan2022PRC}: 
For the particle-hole channel, the relativistic density functional PC-PK1 \cite{Zhao2010PRC} is employed, which can provide one of
the best density-functional descriptions for nuclear masses \cite{Zhao2012PRCmass,Zhang2014FoP,Lu2015PRC,Zhang2021PRC}. 
For the particle-particle channel, a density-dependent zero-range pairing force (4) is adopted, with the pairing strength $V_0 = -325 ~ \mathrm{MeV ~ fm}^3$, the saturation density of nuclear matter $\rho_{\text{sat}} = 0.152 ~ \mathrm{fm}^{-3}$, and a pairing window 100 MeV, which are fixed by reproducing the odd-even mass difference data for calcium and lead isotopes \cite{Zhang2020PRC}. 
The box size is $R_{\text{box}} = 20$ fm and the mesh size is $\Delta r=0.1$ fm. 
For the Dirac Woods-Saxon basis, the angular momentum cutoff is $J_{\max} = 23/2~\hbar$, the energy cutoff is $E_{\text{cut}}^+ = 300$ MeV, and the number of the basis states in the Dirac sea is the same as that in the Fermi sea \cite{Zhou2003PRC}. 
The Legendre expansion truncation for potentials and densities in Eq.~(5) is $\lambda_{\max} = 6$, 8 and 10 for the nuclei with $8 \leqslant Z \leqslant 70$, $71\leqslant Z \leqslant 100$ and $101 \leqslant Z \leqslant 120$, respectively.  
For an odd-$A$ nucleus, the blocking effect is considered with the equal filling approximation \cite{Li2012CPL,Pan2022PRC}. 
The above numerical conditions have been examined and determined in Refs.~\cite{Pan2019IJMPE,Zhang2020PRC,Pan2022PRC}. 

\subsection{ Classical $r$-process }
In this work, a site-independent $r$-process model, i.e., the so-called classical $r$-process model, is employed to specifically study the effect of nuclear mass on the $r$-process simulation. 
Classical $r$-process model can be regarded as a simplification of the dynamical $r$-process model, and has been successfully employed in describing $r$-process patterns of both the solar system and metal-poor stars \cite{Kratz1993APJ,Kratz2007APJ,Sun2008PRC,Xu2013PRC}. 
Nevertheless, it should be noted that the real neutron freeze-out after the equilibrium between neutron capture and photodisintegration reactions, as well as the fission recycling, are neglected in the present classical $r$-process model. 

In the classical $r$-process model, iron group seed nuclei are irradiated by high-density neutron sources with a high temperature $T \gtrsim 1.5~\mathrm{GK}$. 
The $r$-process abundances are obtained by the superposition of abundances from the simulations in 16 different neutron flows with neutron densities in the range of $10^{20}-10^{27.5} \mathrm{~cm}^{-3}$. 
The weight $\omega$ and the irradiation time $\tau$ of each neutron flow follow exponential relations on neutron density $n_n$ \cite{Kratz1993APJ,Chen1995PLB}: 
\begin{equation}
\begin{aligned}
& \tau\left(n_n\right)=a \times n_n^b, \\
& \omega\left(n_n\right)=c \times n_n^d.
\end{aligned}
\end{equation}
The parameters $a, b, c$, and $d$ can be determined from a least-square fit to the solar $r$-process abundances.

In the astrophysical environments with high-temperature $T \gtrsim 1.5~\mathrm{GK}$ and high neutron density $n_n \gtrsim 10^{20} \mathrm{~cm}^{-3}$, the equilibrium between neutron capture and photodisintegration reactions can be achieved, and the abundance ratios of neighboring isotopes on an isotopic chain can be obtained by the Saha equation \cite{Cowan1991PR,Qian2003PPNP,Arnould2007PR}
\begin{equation}
\label{saha}
	\begin{aligned}
\frac{Y(Z, A+1)}{Y(Z, A)}= & n_n\left(\frac{2 \pi \hbar^2}{m_u k T}\right)^{3 / 2} \frac{G(Z, A+1)}{2 G(Z, A)} \\
& \times \exp \left(\frac{A+1}{A}\right)^{3 / 2} \left[\frac{S_n(Z, A+1)}{k T}\right],
	\end{aligned}
\end{equation}
where $Y(Z, A), S_n(Z, A)$, and $G(Z, A)$ are, respectively, the abundance, one-neutron separation energy, and partition function of nuclide $(Z, A)$, and $\hbar, k$, and $m_u$ are the Planck constant, Boltzmann constant, and atomic mass unit, respectively. 
Note that the neutron separation energy $S_n$ deduced from nuclear masses appears in the exponential, suggesting the importance of nuclear masses in the equilibrium.

The abundance flow from one isotopic chain to the next is governed by $\beta$ decays and can be expressed by a set of differential equations
\begin{equation}
\label{beta}
	\begin{aligned}
	\frac{d Y(Z)}{d t}= &  Y(Z-1) \sum_A P(Z-1, A) \lambda_\beta^{Z-1, A} \\
	&  -Y(Z) \sum_A P(Z, A) \lambda_\beta^{Z, A},
	\end{aligned}
\end{equation}
where $\lambda_\beta^{Z, A}$ is the $\beta$ decay rate of the nucleus $(Z,A)$, $Y(Z)=\sum_A Y(Z,A)=\sum_A P(Z, A) Y(Z)$ is the total abundance of each isotopic chain, and $P(Z,A)$ is the individual population coefficients obtained from the equilibrium condition in Eq.~\eqref{saha}.
By using Eqs.~\eqref{saha} and \eqref{beta}, the abundance of each isotope can be determined. After the neutrons freeze-out, the unstable isotopes on the neutron-rich side will proceed to the stable isotopes mainly via $\beta$ decays, and the final abundances are obtained.


\section{ Pseudo {DRHBc} mass table } \label{Smass}

The even-even and even-$Z$ odd-$N$ parts of the DRHBc mass table have been completed, with the numerical details and results summarized in Refs.~\cite{Zhang2022ADNDT,Guo2024ADNDT}, respectively. 
Since the odd-$Z$ part of the DRHBc mass table is not yet available, a proper estimation for the masses of odd-$Z$ nuclei would provide preliminary insights into the knowledge of the whole nuclear landscape and the applications in the $r$-process simulation. 
In Section \ref{iiia}, the methodology of estimating the masses of odd nuclei based on the masses and pairing gaps of even nuclei is shown. 
In Section \ref{iiib}, the examination of the methodology is performed by comparing the estimated masses of even-$Z$ odd-$N$ (even-odd for short) nuclei with the corresponding results in the available DRHBc mass table, and the influence of different pairing gaps is discussed. 
In Section \ref{iiic}, by estimating the masses of odd-$Z$ nuclei based on the available DRHBc results of even-$Z$ nuclei, a pseudo DRHBc mass table is developed and its precision on mass descriptions is evaluated. 

\subsection{Mass estimation for odd nuclei} \label{iiia}

The mass of an odd nucleus can be estimated based on the properties of its neighboring even nuclei \cite{Kortelainen2010PRC,Olsen2013PRL}. 
The binding energy $E_b$ of a nucleus can be expressed in terms of the combinations of its neighbors' quantities, for example, 
 \begin{align}
     E_b (Z,N) = & \frac{1}{2} [E_b(Z,N+1) + E_b(Z,N-1)] \notag \\
     & + \frac{(-1)^N}{2} [\delta_n(Z,N+1)+\delta_n(Z,N-1)], \label{EbN} \\
     \text{or }
     E_b (Z,N) = & \frac{1}{2} [E_b(Z+1,N) + E_b(Z-1,N)] \notag \\
     & + \frac{(-1)^Z}{2} [\delta_p(Z+1,N)+\delta_p(Z-1,N)], \label{EbZ}
 \end{align}
where $\delta_n$ and $\delta_p$ are the three-point odd-even mass differences for neutron and proton, respectively, defined as
 \begin{align}
 	\begin{aligned}
     \delta_n (Z,N) & = \frac{(-1)^N}{2} [2E_b(Z,N) - E_b(Z,N+1) \\
     & \qquad  - E_b(Z,N-1)] , 
 	\end{aligned}\\
 	\begin{aligned}
     \delta_p (Z,N) & = \frac{(-1)^Z}{2} [2E_b(Z,N) - E_b(Z+1,N) \\
     & \qquad - E_b(Z-1,N)] . 
 	\end{aligned}
 \end{align}
The odd-even mass difference is often approximated by the microscopic average pairing gap in Eq.~\eqref{eDel} \cite{Dobaczewski1984NPA,Duguet2001PRC_2},
 \begin{equation}
     \delta_n (Z,N) \approx \Delta_{n}(Z,N),~~~~\delta_p (Z,N) \approx \Delta_{p}(Z,N). 
 \end{equation}
With this approximation, the $E_b$ of an odd nucleus can be obtained based on the $E_b$ and $\Delta_{n/p}$ of its neighboring even nuclei, where the latter ones have already been provided in a self-consistent and microscopic manner from the available part of the DRHBc mass table \cite{Zhang2022ADNDT,Guo2024ADNDT}. 
Such an interpolation treatment on odd nuclei is simple, but contains the nuclear structure information from microscopic calculations to a certain extent, and thus ensures that the magnitude of pairing correlations is correct \cite{Kortelainen2010PRC}. 

\subsection{Examination of the mass estimation: From even-even to even-odd nuclei} \label{iiib}

\begin{figure}
\centering
\resizebox{\linewidth}{!}{\includegraphics{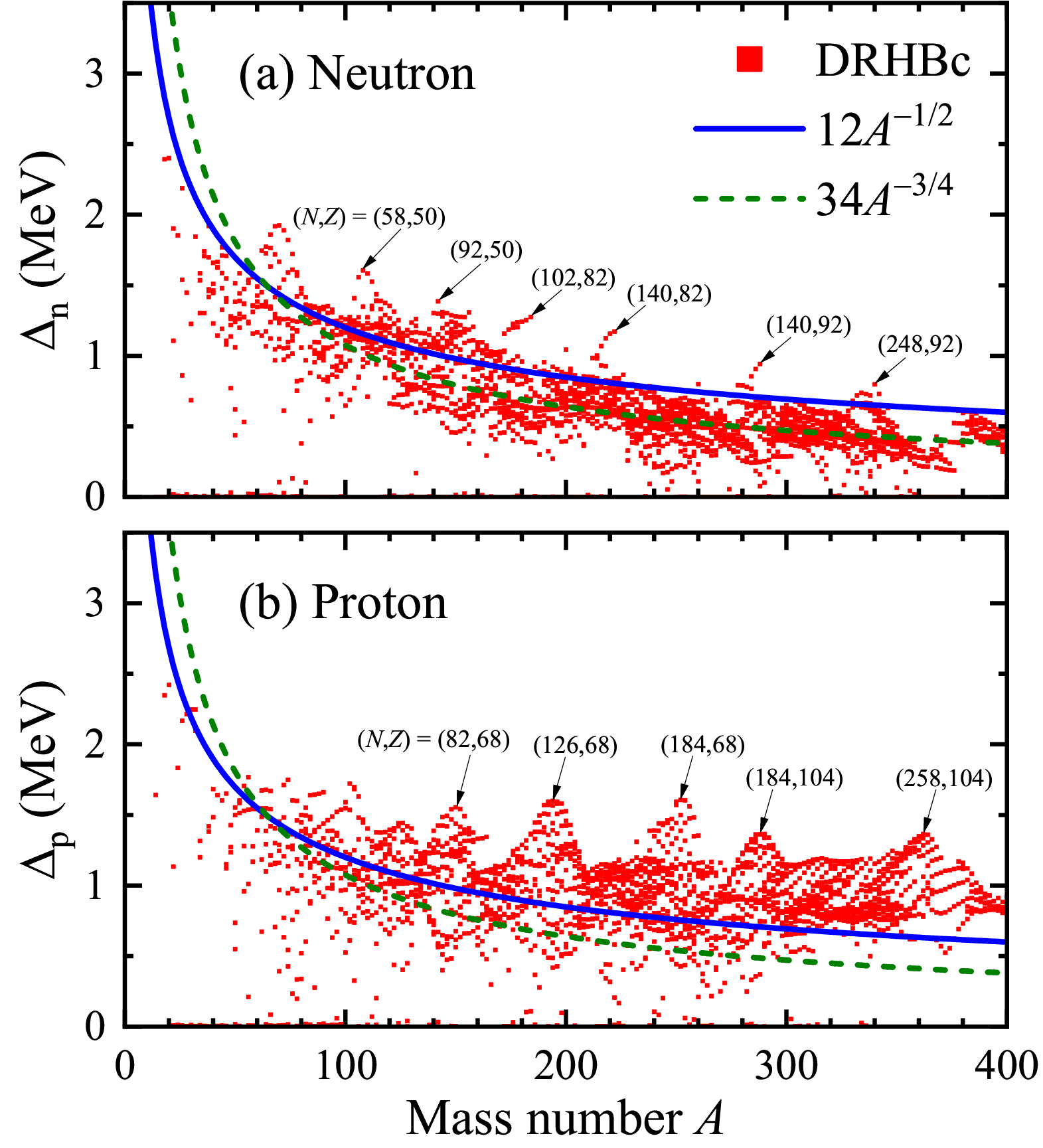}}
\caption{Average pairing gaps of even-even nuclei with $8\leqslant Z \leqslant 120$ in DRHBc theory for (a) neutron and (b) proton, in comparison with the empirical pairing gaps $\Delta=12A^{-1/2}$ and $\Delta=34A^{-3/4}$ \cite{Ring1980NMBP,Bohr1998book}, respectively.}
\label{mass_Delta}
\end{figure}

In Fig.~\ref{mass_Delta}, the average pairing gaps calculated by Eq.~(8) for the even-even nuclei with $8\leqslant Z \leqslant 120$ in DRHBc mass table are shown for neutron and proton, respectively. 
The empirical pairing gaps, $\Delta=12A^{-1/2} $ from the experimental odd-even mass staggering, and $\Delta=34A^{-3/4}$ from the semi-empirical mass formula of the liquid drop model \cite{Ring1980NMBP,Bohr1998book}, are also shown for comparison. 

For the neutron pairing gap $\Delta_n$ in Fig.~\ref{mass_Delta}(a), generally, the DRHBc result decreases with the increase of $A$. 
The DRHBc results are in rough agreement with both empirical curves $12A^{-1/2}$ and $34A^{-3/4}$.  
Several peaks occur with the proton numbers at shell closures $Z=20, 50$ and $82$, or at the pseudo shell $92$ \cite{Geng2006CPL,Zhang2022ADNDT}, while their neutron numbers mainly locate in the middle of the shells. 
This is attributed to the spherical shape at the proton shell closure and the high single-neutron level density in the middle of the neutron shell, leading to the strong pairing effect for neutron. 

For the proton pairing gap $\Delta_p$ in Fig.~\ref{mass_Delta}(b), the DRHBc result shows a decreasing trend at $A<200$, while no significant decreasing or increasing trend occurs at $A>200$. 
This is because at a large $A$, the increase of $A$ mainly comes from the increase of $N$, instead of $Z$. 
Several peaks occur with the neutron numbers at shell closures $N=82, 126, 184$ and $258$, similar to the behavior of $\Delta_n$ in Fig.~\ref{mass_Delta}(a). 
Comparing Figs. \ref{mass_Delta} (a) and (b), it is found that $\Delta_p$ tends to locate above the empirical curves, while $\Delta_n$ tends to locate below the empirical ones at large $A$. 
Such difference originates from the general trend in the DRHBc calculations that pairing gap weakens as the number of the corresponding nucleon increases, i.e., $\Delta_n$ mainly depends on $N$ and $\Delta_p$ mainly depends on $Z$. 
Since $N>Z$ in most nuclei, as $A$ increases, $N$ grows faster than $Z$ in a general trend. 
This leads to a more rapid decreasing trend in $\Delta_n$ than in $\Delta_p$ for the DRHBc results and results in $\Delta_n$ lying below the empirical curves at large $A$, as the empirical ones only depend on $A$.

\begin{figure}
\centering
\resizebox{\linewidth}{!}{\includegraphics{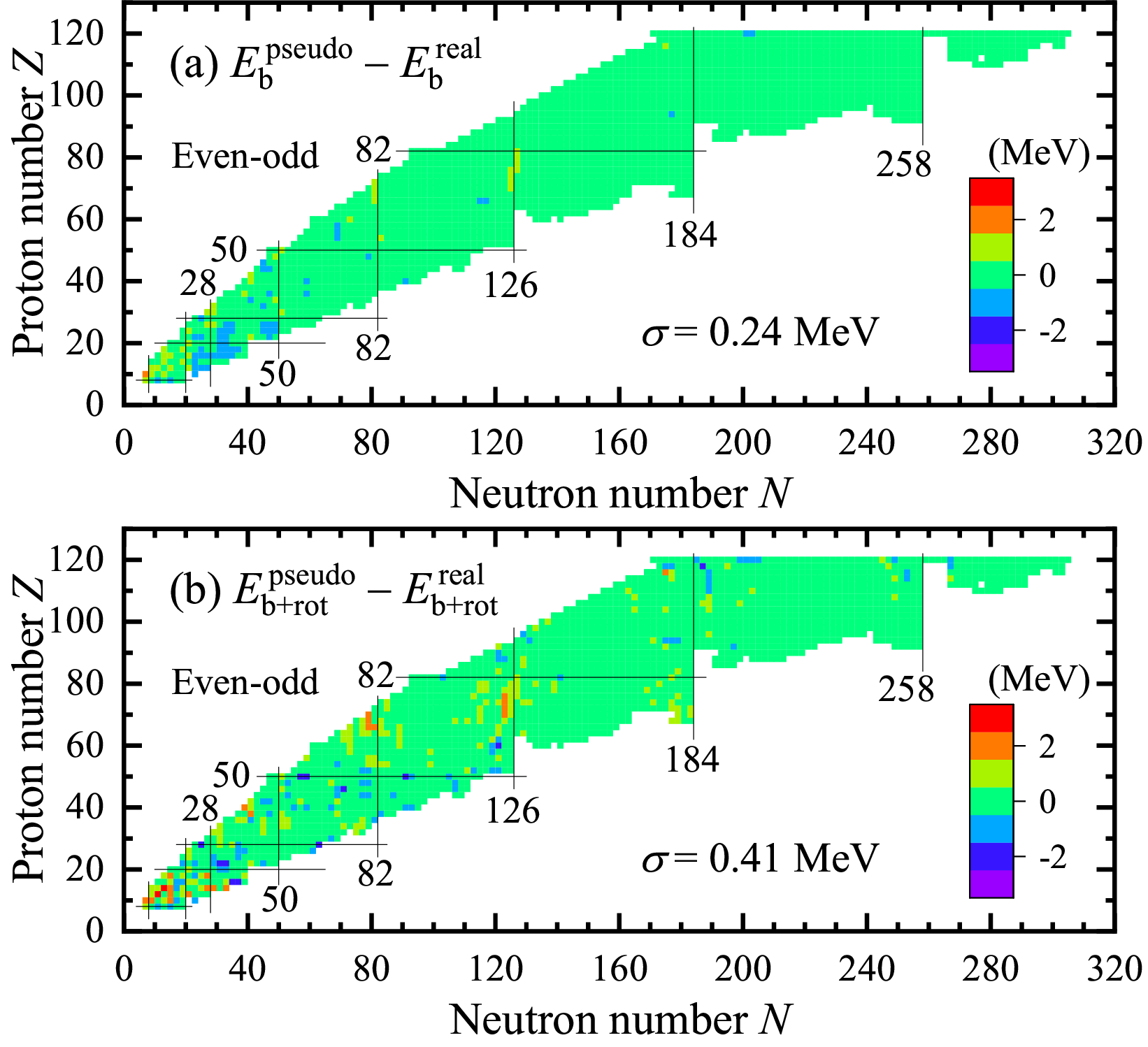}}
\caption{The differences between the pseudo and real DRHBc results for the binding energies of even-odd nuclei with $8 \leqslant Z \leqslant 120$ scaled by colors. 
    (a) The rotational correction is not considered. 
    (b) The rotational correction is included. }
\label{mass_exam}
\end{figure}

To validate the methodology in Section \ref{iiia}, by taking all the even-odd nuclei in the nuclear chart with $8 \leqslant Z \leqslant 120$ as examples, their binding energies $E_b$ are estimated based on the $E_b$ and $\Delta_n$ of the even-even nuclei, and compared with the available results in the DRHBc mass table \cite{Guo2024ADNDT}. 
Since the estimated results are not obtained from the real DRHBc calculations, for convenience, they are labeled ``pseudo'' DRHBc results in this work. 
Figure \ref{mass_exam} shows the differences between the pseudo binding energies and the real DRHBc results from self-consistent blocking calculations (labeled as ``real'') for even-odd nuclei.
In Fig.~\ref{mass_exam}(a), where the rotational correction is not considered, for most nuclei the differences between the pseudo and real DRHBc results are less than 0.5 MeV, with the root-mean-square (rms) difference being 0.24 MeV, showing excellent agreement between these two treatments. 
In comparison, the rms deviation of real DRHBc binding energies (w/o rotational correction) from experimental data for even-$Z$ nuclei is 2.56 MeV \cite{Guo2024ADNDT}. 
The difference between the two theoretical treatments is dramatically smaller than the deviation from the data. 
This indicates that the pseudo DRHBc binding energies can well reproduce the real DRHBc results. 

For the density functional PC-PK1 adopted here, the rotational correction has been shown to play an important role in improving the binding energy description for deformed nuclei \cite{Zhao2010PRC,Zhang2020PRC,Pan2022PRC}. 
Therefore, it is necessary to compare $E_{\text{b}}^{\text{pseudo}}$ and $E_{\text{b}}^{\text{real}}$ after incorporating the rotational correction energy $E_{\text{rot}}$, and the corresponding results are shown in Fig.~\ref{mass_exam}(b). 
The differences between these two treatments are still less than 0.5 MeV for most nuclei, while we also noticed that some nuclei exhibit larger differences than those in Fig.~\ref{mass_exam}(a). 
For example, for the nuclei with $N=121,123$, $Z\approx 70$, as well as those with $N=77,79$, $Z\approx 67$, their $E_{\text{b+rot}}^{\text{pseudo}} - E_{\text{b+rot}}^{\text{real}}$ are larger than 1.5 MeV. 
These larger differences mainly correspond to the abrupt changes of nuclear shape, especially those near magic numbers, whose deformations suddenly decrease to (near-)zero. 
In this case, the cranking approximation is not suitable for $E_{\text{rot}}$ in Eq.~\eqref{Erot}.
Considering that the $E_{\text{b+rot}}^{\text{pseudo}}$ for an even-odd nucleus in Fig.~\ref{mass_exam}(b) is interpolated based on the $E_{\text{b+rot}}^{\text{real}}$ of its neighboring even-even nuclei, the pseudo result may significantly deviate from the real result when the $\beta_2$ values of neighboring even-even nuclei straddle 0.05. 
Although some nuclei show larger differences, the rms difference in Fig.~\ref{mass_exam}(b) is 0.41 MeV, which is slightly larger than that without $E_{\text{rot}}$ in Fig.~\ref{mass_exam}(a), but still much smaller than the rms deviation of $E_{\text{b+rot}}^{\text{real}}$ from the data, 1.43 MeV. 
Therefore, after including the rotational correction, the pseudo and real DRHBc binding energies are still in good agreement for most nuclei, and their differences are not expected to substantially influence the discussions on physics. 

\begin{figure}
\centering
\resizebox{\linewidth}{!}{\includegraphics{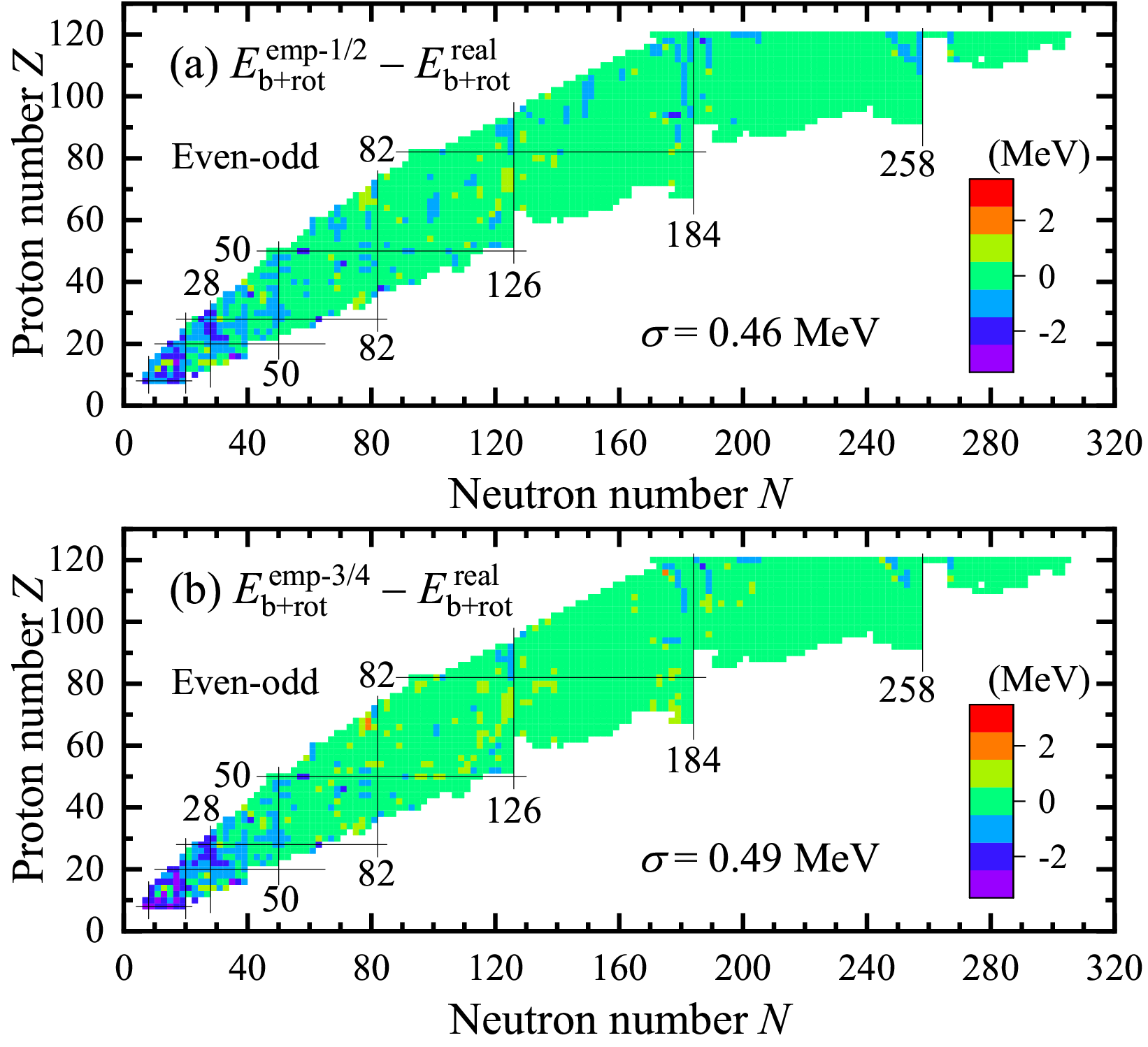}}
\caption{ The same as Fig.~\ref{mass_exam}, but the pairing gaps are estimated with the empirical formulas (a) $\Delta=12A^{-1/2}$ (``emp-1/2'' for short) and (b) $\Delta=34A^{-3/4}$ (``emp-3/4'' for short). The rotational correction is included. }
\label{mass_emp}
\end{figure}

To further examine the methodology, based on the masses of even-even nuclei and the empirical pairing gaps $\Delta = 12 A^{-1/2}$ and $\Delta=34 A^{-3/4}$, the binding energies of even-odd nuclei are also estimated, respectively, for comparisons. 
In Fig.~\ref{mass_emp}, the estimated binding energies are compared with the real DRHBc results. 

For the estimation based on the empirical pairing gap $\Delta= 12 A^{-1/2}$ shown in Fig.~\ref{mass_emp}(a), the rms difference is $\sigma=0.46$ MeV, larger than that in Fig.~\ref{mass_exam}(b). 
For the nuclei with $Z<28$, noticeable differences with $E_{\text{b+rot}}^{\text{emp-1/2}} - E_{\text{b+rot}}^{\text{real}} \approx -2 $ MeV are observed. 
Such differences can be understood from Fig.~\ref{mass_Delta}(a), where the $12A^{-1/2}$ curve is higher than most of the DRHBc results when $A$ is small, leading the underestimations of the binding energies according to Eq.~\eqref{EbN}. 
  
For the estimation based on the empirical pairing gap $\Delta= 34 A^{-3/4}$ shown in Fig.~\ref{mass_emp}(b), the rms difference is $\sigma=0.49$ MeV, larger than those in Fig.~\ref{mass_exam}(b) and \ref{mass_emp}(a). 
For the nuclei with $Z<28$, the differences reach $E_{\text{b+rot}}^{\text{emp-3/4}} - E_{\text{b+rot}}^{\text{real}} \approx -3$ MeV, larger than those in Fig.~\ref{mass_emp}(a). 
This is attributed to the fact that the $34 A^{-1/2}$ curve is higher than the $12 A^{-1/2}$ curve when $A$ is small in Fig.~\ref{mass_Delta}(a), leading to more underestimated binding energies according to Eq.~\eqref{EbN}.

According to above discussions, the pseudo DRHBc results nicely reproduce the binding energies in the real DRHBc mass table for even-odd nuclei. 
The comparisons with the results based on the empirical pairing gaps show that the global behavior of the binding energy is not very sensitive to the details in the evolution of pairing gaps.

\subsection{The pseudo mass table: From even-$Z$ to odd-$Z$ nuclei} \label{iiic}

Based on the DRHBc mass table for even-$Z$ nuclei \cite{Guo2024ADNDT} and using Eqs.~\eqref{EbN} and \eqref{EbZ} ($\delta_{n/p}$ are approximated by $\Delta_{n/p}$), the binding energies of odd-$Z$ nuclei with $8 \leqslant Z \leqslant 120$ are estimated. 
Combining the pseudo binding energies of odd-$Z$ nuclei and the real DRHBc results of even-$Z$ nuclei available, a pseudo DRHBc mass table is obtained. 
For convenience, here the corresponding binding energy value is labeled as $E_{\text{b}}^{\text{pseudo}}$, which includes both pseudo results for odd-$Z$ nuclei and real DRHBc results for even-$Z$ nuclei. 
9480 bound nuclei are obtained in total, where 2584 (27.3\%) are even-even, 2245 (23.7\%) are even-odd, 2513 (26.5\%) are odd-even, and 2138 (22.6\%) are odd-odd. 

\begin{figure}
    \centering
    \resizebox{\linewidth}{!}{\includegraphics{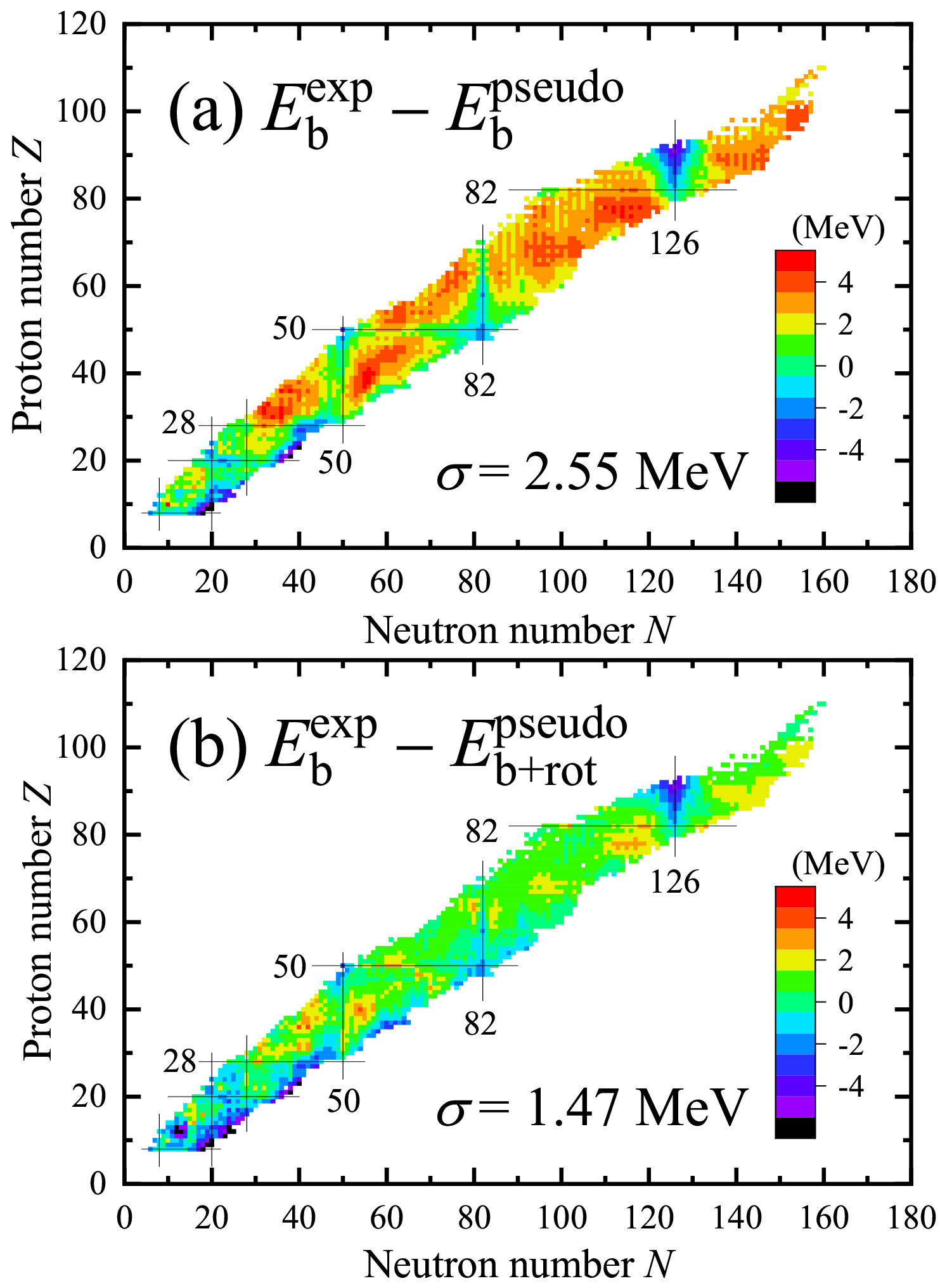}}
    \caption{ The differences between the pseudo DRHBc binding energies and the available data \cite{Wang2021CPC} for the nuclei with $8 \leqslant Z \leqslant 120$ scaled by colors. 
    The details about the pseudo DRHBc results can be found in the text. 
    (a) The rotational correction is not considered. 
    (b) The rotational correction is included.  
    }
    \label{mass_Eb}
\end{figure}

Among these 9480 bound nuclei, the masses of 2413 nuclei have been measured experimentally \cite{Wang2021CPC}. 
Figure \ref{mass_Eb} shows the binding energy deviations $E_b^{\text{exp}} - E_b^{\text{pseudo}}$ for these measured nuclei. 
The results without $E_{\text{rot}}$ and with $E_{\text{rot}}$ are shown in panels (a) and (b), respectively. 
In Fig.~\ref{mass_Eb}(a), the deviation is relatively small near magic numbers, while it becomes larger when getting far away from magic numbers, which is related to the increase of deformation. 
The rms deviation of predicted binding energies from data is $\sigma=2.55$ MeV. 
In Fig.~\ref{mass_Eb}(b), after including the $E_{\text{rot}}$, the binding energy description is significantly improved for most nuclei, especially those with both neutron and proton numbers far away from magic numbers. 
Accordingly, the rms deviation reduces to $\sigma=1.47$ MeV. 
As a comparison, the corresponding rms deviation for real DRHBc binding energies for only even-$Z$ nuclei is $\sigma=2.56$ MeV without $E_{\text{rot}}$, and $\sigma=1.43$ MeV with $E_{\text{rot}}$ \cite{Guo2024ADNDT}. 
The rms deviations in Fig.~\ref{mass_Eb} are very close to these two values, respectively, indicating that our pseudo binding energies of odd-$Z$ nuclei reach almost the same accuracy as the real DRHBc results of even-$Z$ nuclei. 

\begin{figure}
    \centering
    \resizebox{\linewidth}{!}{\includegraphics{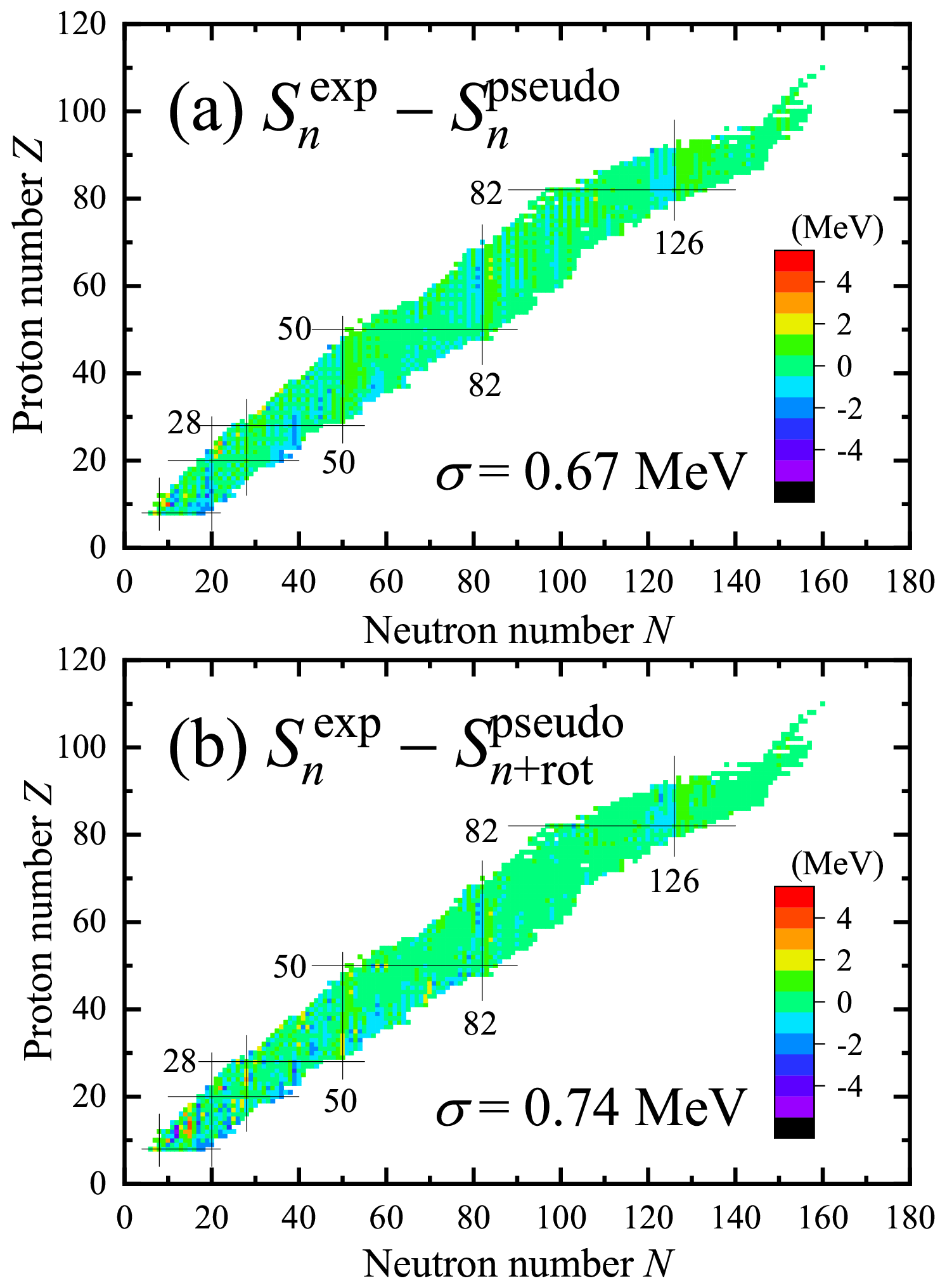}}
    \caption{The differences between the pseudo DRHBc results of one-neutron separation energy and the available data \cite{Wang2021CPC} for the nuclei with $8 \leqslant Z \leqslant 120$ scaled by colors. 
    (a) The rotational correction is not considered. 
    (b) The rotational correction is included. }
    \label{mass_Sn}
\end{figure}

The one-neutron separation energy
\begin{equation}
    S_n(Z,N) = E_b(Z,N) - E_b(Z,N-1) 
\end{equation}
is intricately connected to the characteristics of the astrophysical $r$-process. 
Utilizing the $E_b^{\text{pseudo}}$ in Fig.~\ref{mass_Eb}(a), the $S_n^{\text{pseudo}}$ without rotational correction is calculated and its deviation from 2299 available data \cite{Wang2021CPC} is shown in Fig.~\ref{mass_Sn}(a). 
The deviations for most nuclei are within 1 MeV, and the rms deviation $\sigma=0.67$ MeV. 
%
Figure \ref{mass_Sn}(b) shows the deviation of $S_n$ after including the rotational correction energy. 
It is found that, generally, $E_{\text{rot}}$ improves the description of $S_n$ at $Z>50$, but for light nuclei the deviation is increased, resulting in the rms deviation $\sigma=0.74$ MeV, slightly larger than the case without rotational correction in panel (a). 
The result that $E_{\text{rot}}$ does not improve the $\sigma$ of $S_n$ has been noticed in the calculations on even-$Z$ nuclei \cite{Guo2024ADNDT}, and is attributed to the limit of cranking approximation, which has already been discussed in Fig.~\ref{mass_exam}(b). 
We also noticed that the deviations in some odd-$Z$ nuclei are different from those of their even-$Z$ neighbors, forming odd-even staggering with respect to proton number in Fig.~\ref{mass_Sn}. 
This means that these pseudo results do not fully captures the odd-even effects.
We expect that a fully self-consistent treatment in the complete DRHBc mass table in the future would eliminate this staggering in the deviation of $S_n$. 

\begin{table*}
    \caption{ The rms deviations $\sigma$ for binding energies ($E_b$), one- and two-neutron separation energies ($S_n,S_{2n}$), as well as one- and two-proton separation energies ($S_p, S_{2p}$) in the pseudo DRHBc mass table with respect to the available data \cite{Wang2021CPC} in the unit of MeV. 
    For comparison, the corresponding deviations for even-$Z$ and odd-$Z$ nuclei are also listed separately. }
    \centering
    \begin{tabular}{c|c|ccccc}
    \hline 
    Range  & Theory & $\sigma(E_b)$ & $\sigma(S_n)$ & $\sigma(S_{2n})$ & $\sigma(S_p)$ & $\sigma(S_{2p})$ \\
     \hline 
     All & Pseudo + real DRHBc (w/o $E_{\text{rot}}$) & 2.55 & 0.67 & 0.93 & 0.66 & 0.93 \\
     All & Pseudo + real DRHBc (w/  $E_{\text{rot}}$) & 1.47 & 0.74 & 0.96 & 0.71 & 0.98 \\
     \hline 
     Even-$Z$ & Real DRHBc (w/o $E_{\text{rot}}$) & 2.56 & 0.75 & 0.95 & - & 0.93 \\
     Even-$Z$ & Real DRHBc (w/  $E_{\text{rot}}$) & 1.43 & 0.77 & 0.99 & - & 1.05 \\
     \hline 
     Odd-$Z$ & Pseudo DRHBc (w/o $E_{\text{rot}}$) & 2.54 & 0.58 & 0.89 & - & 0.93 \\
     Odd-$Z$ & Pseudo DRHBc (w/  $E_{\text{rot}}$) & 1.50 & 0.70 & 0.94 & - & 0.90 \\
     \hline 
    \end{tabular}
    \label{t_sigma}
\end{table*}

Similar to the discussions on Figs.~\ref{mass_Eb} and \ref{mass_Sn}, the rms deviations of other predicted quantities, including the two-neutron separation energy $S_{2n}$ and one- and two-proton separation energies $S_p, S_{2p}$, in the pseudo DRHBc mass table are calculated and summarized in Table \ref{t_sigma}. 
It is noted that by introducing $E_{\text{rot}}$, the description of $E_b$ is significantly improved, while for separation energies the corresponding $\sigma$ values slightly increase, due to the cranking approximation as mentioned in the above discussions. 
For comparison, the corresponding results for even-$Z$ and odd-$Z$ nuclei are also listed separately in Table \ref{t_sigma}. 
It is found that odd-$Z$ nuclei have similar $\sigma$ values to even-$Z$ ones for both binding energies and separation energies, showing the consistency between the pseudo and real DRHBc results. 

In addition, based on the empirical pairing gaps $\Delta = 12A^{-1/2}$ and $\Delta=34A^{-3/4}$, the masses of odd-$Z$ nuclei are estimated in the same manner. 
The corresponding $\sigma(E_b)$ (w/o $E_{\mathrm{rot}}$) are 2.66 MeV and 2.61 MeV,  respectively. 
The corresponding $\sigma(E_b)$ (w/  $E_{\mathrm{rot}}$) are 1.47 MeV and 1.43 MeV, respectively.  
These values are close to the $\sigma(E_b)$ in Table \ref{t_sigma}. 
The impacts on $\sigma(E_b)$ due to the differences among these three sets of pairing gaps are less than 0.1 MeV. 
Such small differences are consistent with the conclusion in Section \ref{iiib} that the global behavior of $E_b$ is not very sensitive to the details in the evolution of pairing gaps. 

In conclusion, in this Section, based on the available DRHBc results \cite{Guo2024ADNDT} for even-$Z$ nuclei, the binding energies of odd-$Z$ nuclei are estimated. 
A complete pseudo DRHBc mass table is obtained for all nuclei with $8 \leqslant Z \leqslant 120$, with the accuracies in describing nuclear masses and separation energies expected to be close to the real DRHBc results from self-consistent calculations. 

\section{  $R$-process simulation } \label{Srpro}

The obtained pseudo DRHBc mass table in Section \ref{Smass} is applied in the $r$-process simulation, to study the impact of deformation effects in the $r$-process.
The $r$-process simulation is performed using the classical $r$-process model, where nuclear masses are taken from the pseudo DRHBc mass table, if the experimental data~\cite{Wang2021CPC} are not available. 
For the $\beta$-decay rates, the empirical formula~\cite{Zhou2017NPR} using decay energies from the pseudo DRHBc calculations is employed together with experimental data~\cite{Audi2003NPA}. 
The astrophysical trajectory, i.e., the weight $\omega$ and the irradiation time $\tau$ of neutron flows, is determined by fitting the obtained abundances to the solar $r$-process abundances~\cite{Sneden2008ARNPS} at a temperature of $T=1.5 \ \mathrm{GK}$. For comparison, corresponding $r$-process simulation based on the RCHB mass table~\cite{Xia2018ADNDT} is also carried out.

    \begin{figure}
    	\centering
    	\resizebox{\linewidth}{!}{\includegraphics{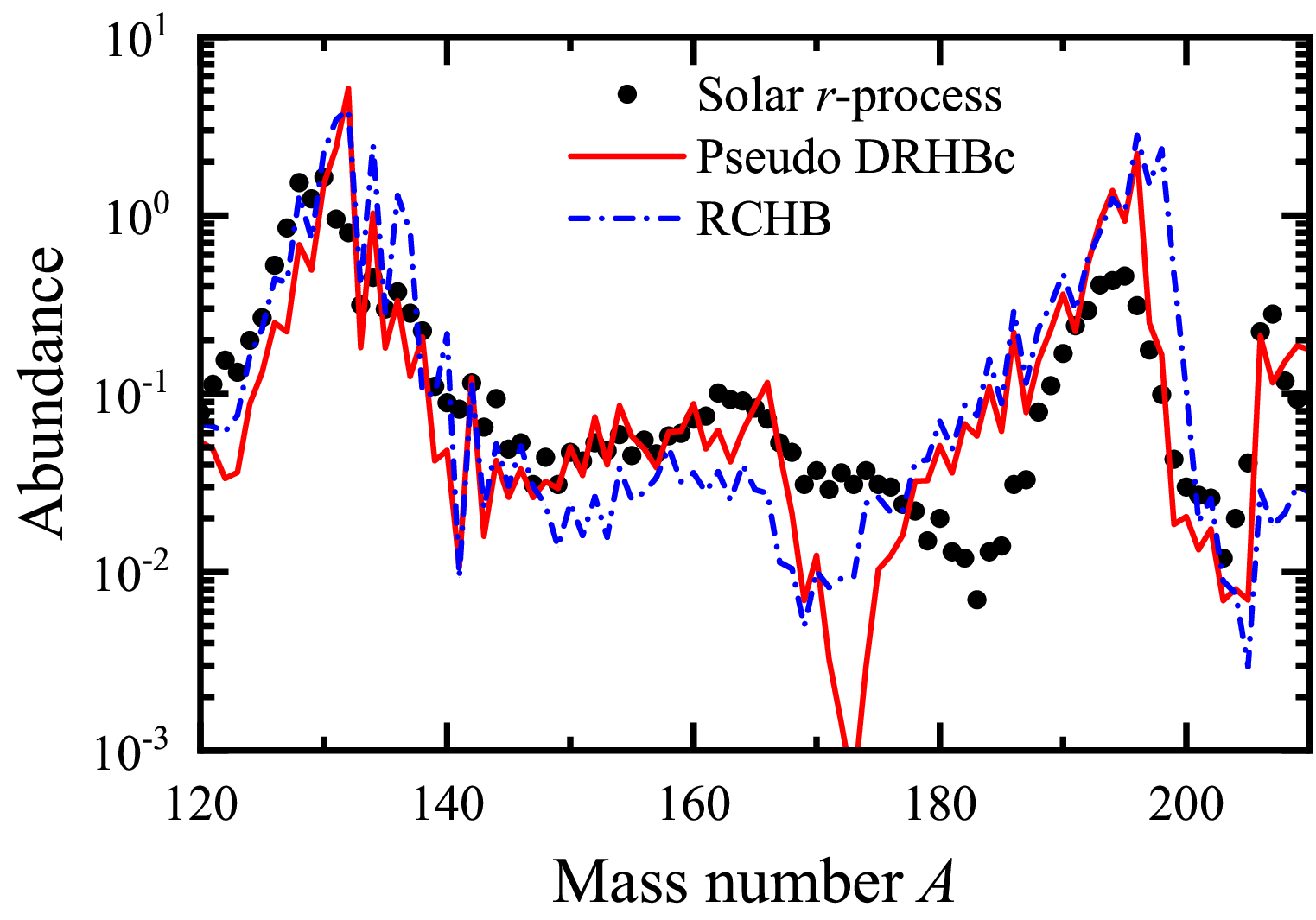}}
    	\caption{The $r$-process abundances from classical $r$-process simulations based on pseudo DRHBc and RCHB mass tables as functions of the mass number $A$. The solar $r$-process abundances \cite{Sneden2008ARNPS} are displayed with black dots.}
        \label{r_abundance}
    \end{figure}
    
The $r$-process abundances with pseudo DRHBc and RCHB mass tables are shown in Fig. \ref{r_abundance}. 
It is found that the simulated abundances with both the RCHB and pseudo DRHBc mass tables exhibit abundance peaks near $A=130,195$ and an abundance trough around $A=170$. 
The $r$-process abundances based on pseudo DRHBc mass table are higher for $A=148-165$ and lower for $A=170-178$ compared to those based on the RCHB mass table. 
This indicates the impact from the deformation effects.  
We note that discrepancies exist between Solar $r$-process abundances and the ones based on pseudo DRHBc or RCHB mass tables. 
The reasons for this can be multifaceted. 
For example, the triaxial and octupole degrees of freedom, as well as beyond-mean-field effects, are not considered in current treatments, which may lead to deviations in nuclear masses. 
Beyond the approximations in symmetry, uncertainties also exist in density functional calculations due to different fitting protocols \cite{Agbemava2014PRC}. 
The approximations in the classical $r$-process simulations, e.g., waiting point approximation and sudden freeze-out approximation, can also contribute to such discrepancies.

    \begin{figure}
        \centering
        \resizebox{\linewidth}{!}{\includegraphics{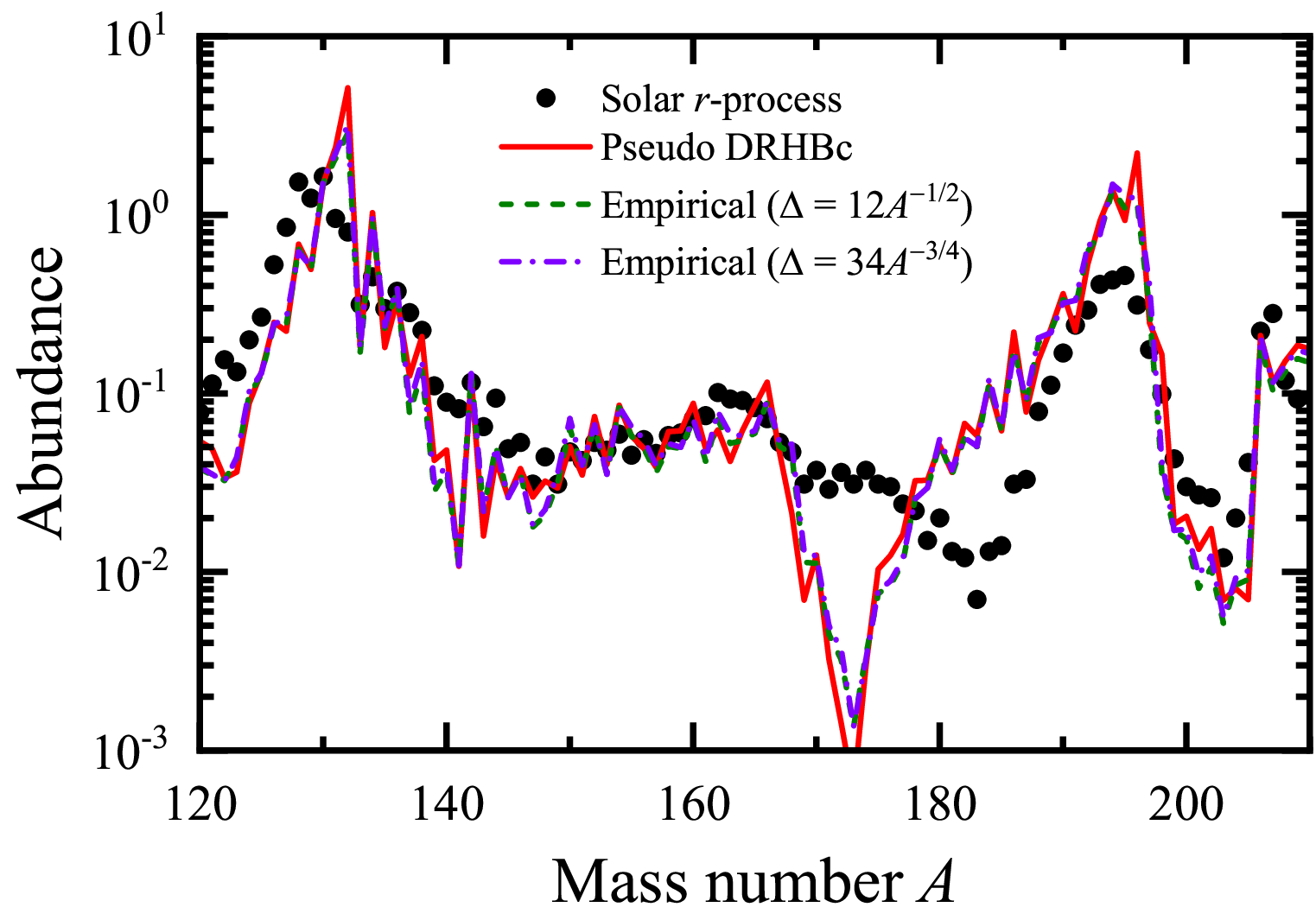}}
        \caption{The $r$-process abundances from classical $r$-process simulations based on the masses of even-$Z$ nuclei from the DRHBc mass table and the masses of odd-$Z$ nuclei from estimations using different pairing gaps, i.e., the microscopic average pairing gaps (pseudo DRHBc), the empirical gaps $\Delta=12A^{-1/2}$, and  $\Delta=34A^{-3/4}$. 
        The solar $r$-process abundances \cite{Sneden2008ARNPS} are displayed with black dots. }
        \label{r_abundance_gap}
    \end{figure}
   
Before proceeding to explore the deformation effect, considering that the input masses of odd-$Z$ nuclei are estimated based on the average pairing gaps in Section \ref{iiic}, it would be important to examine whether $r$-process simulation is sensitive to the choice of pairing gaps. 
In Fig.~\ref{r_abundance_gap}, the $r$-process abundances are shown, with the masses of odd-$Z$ nuclei from estimations using the microscopic average pairing gaps (pseudo DRHBc) and the empirical gaps. 
It is found that the $r$-process abundances from empirical gaps exhibit qualitatively the same behaviors as the pseudo DRHBc results. 
This indicates that these pairing-gap choices are sufficient to support the estimation of masses for nuclei that have not been explicitly calculated.
       
\begin{figure}[htbp]
	\centering
	\resizebox{\linewidth}{!}{\includegraphics{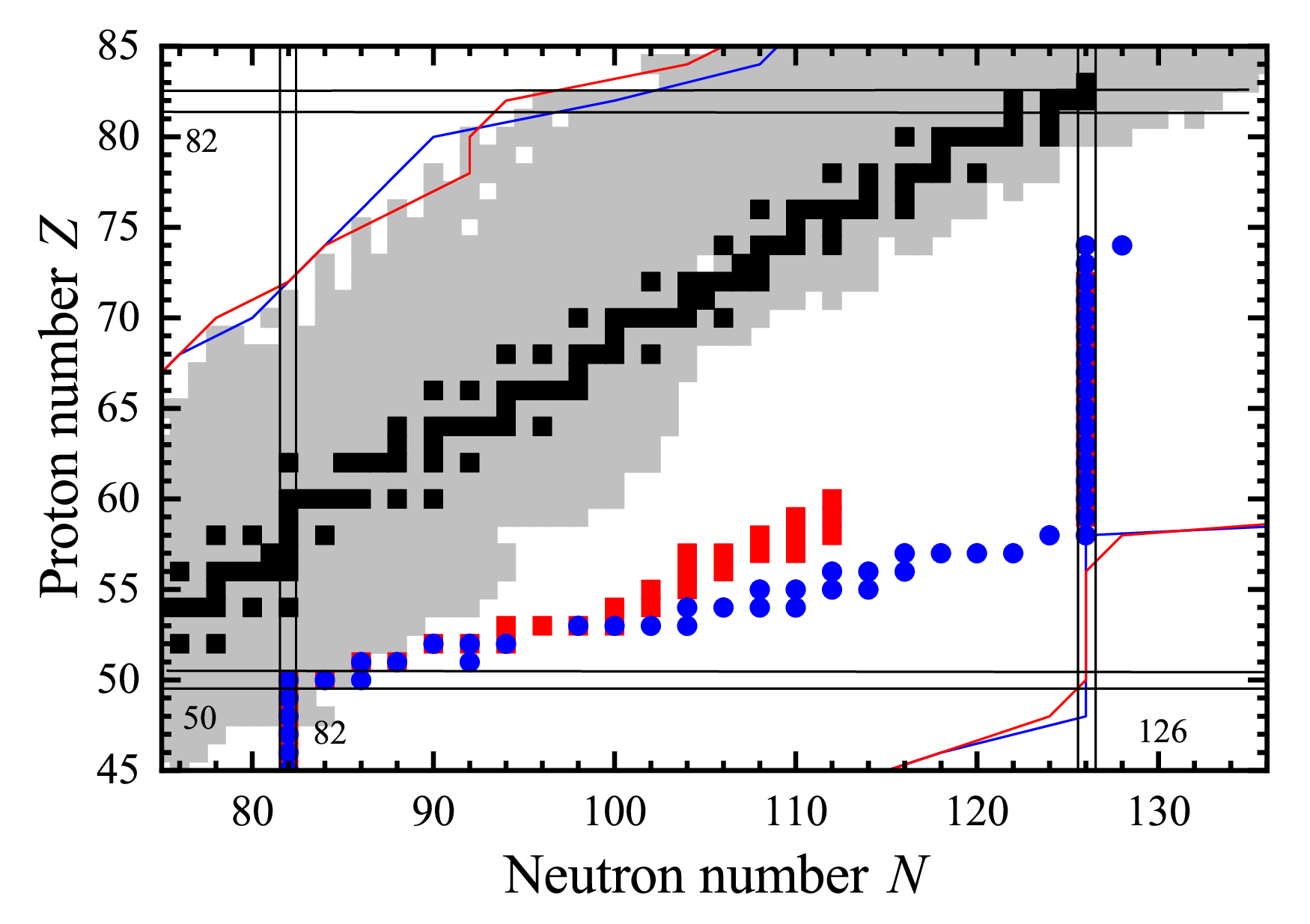}}
	\caption{The $r$-process path for nuclei with $A\sim 130-195$ based on the pseudo DRHBc and RCHB mass tables denoted as red filled squares and blue filled circles, respectively. 
	 Stable nuclei (black filled square), experimentally known nuclei~\cite{Wang2021CPC} (gray filled square), drip lines predicted by the pseudo DRHBc (red line) and RCHB (blue line) mass tables are also displayed.}
    \label{r-path}
\end{figure}

The most important nuclei for the $r$-process are located in the $r$-process path.
The $r$-process paths for nuclei with $A\sim 130-195$ based on the pseudo DRHBc and RCHB mass tables are displayed in Fig. \ref{r-path}. 
It is found that the $r$-process path nuclei around $A=160$ predicted by the pseudo DRHBc mass table are closer to the stability line than the ones predicted by the RCHB mass table. 
Since the beta decay half-lives are generally shorter when the nuclei are moving toward the neutron-rich direction, the decay half-lives of $r$-process path nuclei around $A=160$ based on the pseudo DRHBc mass table are longer than those based on the RCHB mass table.
This leads to increased accumulation of abundances on the path nuclei.
It is therefore understandable that the $r$-process abundances based on the pseudo DRHBc mass table are larger for $A=148-165$ than the ones based on the RCHB mass table. 
On the other hand, it is noted that the $r$-process path predicted by the pseudo DRHBc mass table has a gap around $A = 175$, which leads to a deficiency in the simulated abundances in this region.

\begin{figure}[htbp]
	\centering
	\resizebox{\linewidth}{!}{\includegraphics{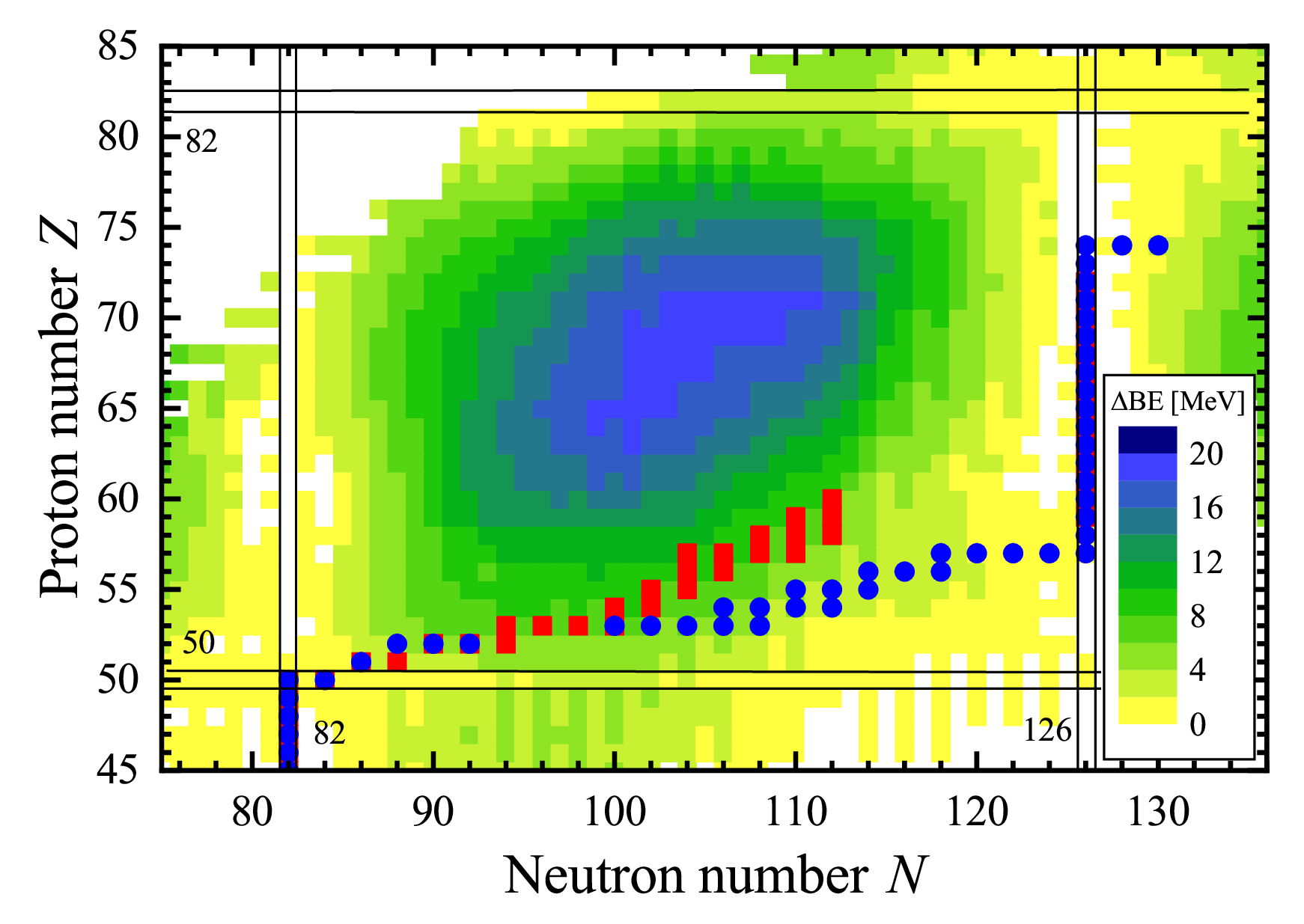}}
	\caption{The binding energy differences for bound nuclei with $82 \le N \le 126$ between the pseudo DRHBc calculations and RCHB calculations scaled by colors.
     The $r$-process path for nuclei with $A\sim 130-195$ based on the pseudo DRHBc and RCHB mass tables denoted as red filled squares and blue filled circles, respectively.  }
    \label{mass_diff}
\end{figure}

\begin{figure}[htbp]
	\centering
	\resizebox{\linewidth}{!}{\includegraphics{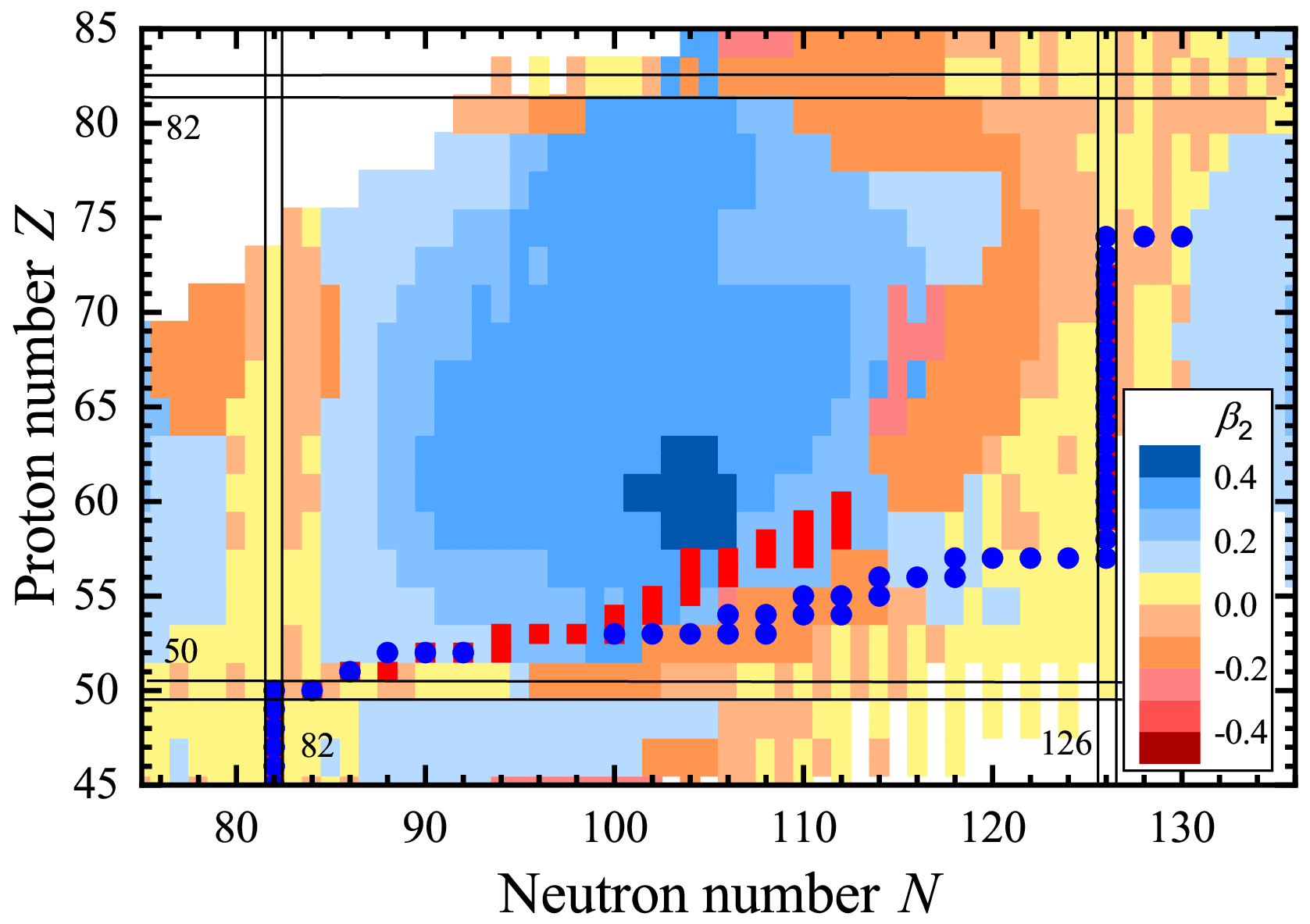}}
	\caption{Quadrupole deformations $\beta_2$ from the DRHBc calculations for bound even-$Z$ nuclei with $82 \le N \le 126$ scaled by colors.}
    \label{def_diff}
\end{figure}

Figure \ref{mass_diff} shows the differences of nuclear binding energy predicted by the pseudo DRHBc and RCHB mass tables. The differences of nuclear binding energy predictions are relatively small around neutron and proton magic numbers, while they become larger in the middle of the shell. 
In the simulation, most $r$-process path nuclei with $A=134-181$ are located in the $N=82-126$ region, where the pseudo DRHBc mass table predicts higher binding energies than the RCHB mass table due to the deformation effect. 
This leads to the differences of $r$-process abundances between these two mass models.
The mass differences are mainly caused by the deformation effects.
Figure \ref{def_diff} shows the quadrupole deformation $\beta_2$ of even-$Z$ nuclei predicted by the DRHBc theory. 
The spherical and near-spherical nuclei are often found around neutron and proton magic numbers as expected, and the well-deformed nuclei often appear in the middle of the shell. 
Meanwhile the prolate-oblate shape changes can be found around $N=120$ for $50\le Z\le82$. 
Notably, the $r$-process path with $A=134-181$ passes through the region for $N=82-126$. 
This indicates that the prediction of $r$-process path with $A=134-181$ is significantly influenced by the deformation effect.

\begin{figure}[htbp]
	\centering
	\resizebox{\linewidth}{!}{\includegraphics{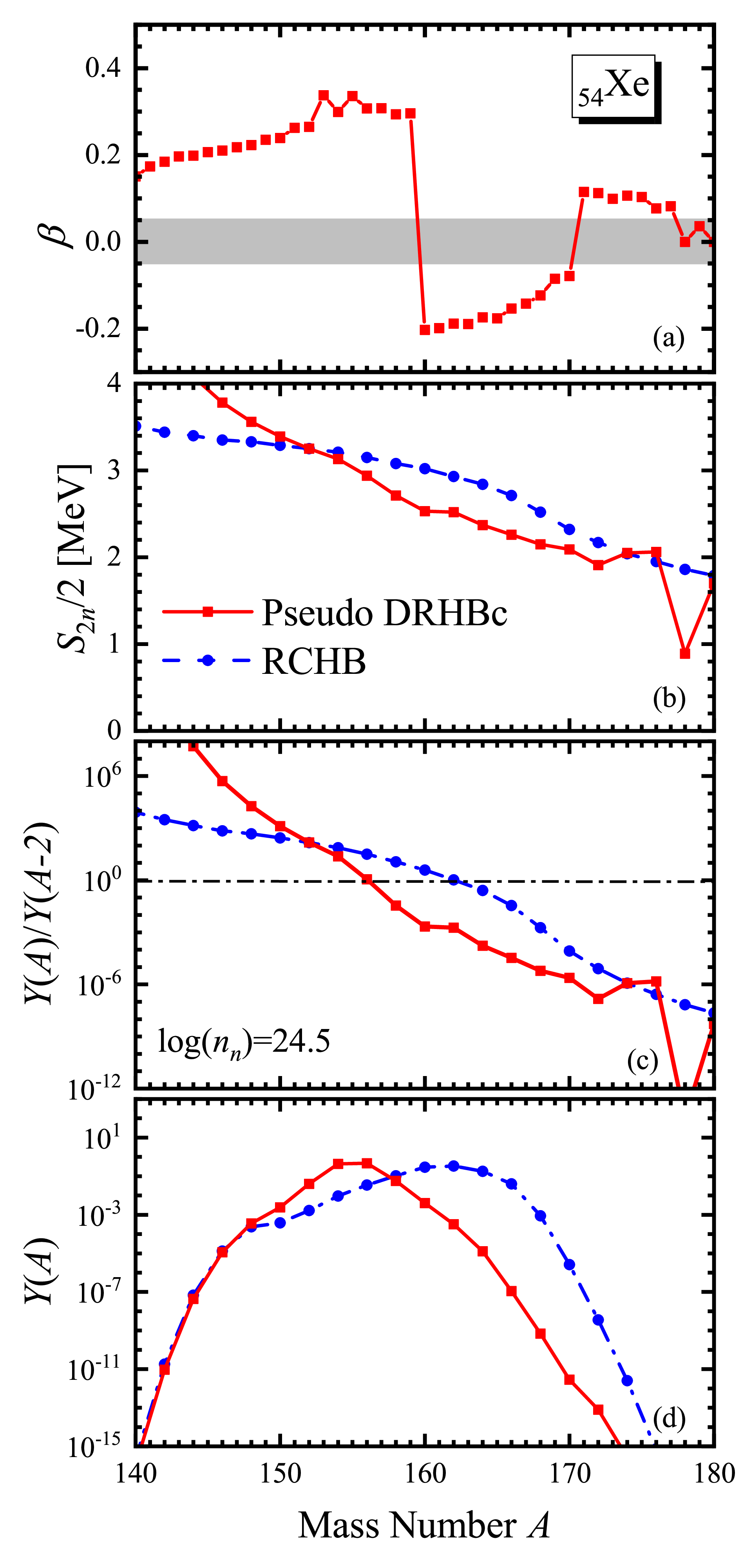}}
	\caption{(a) The nuclear quadrupole deformations $\beta_2$ of xenon isotopes predicted by the DRHBc theory as a function of mass number $A$. (b) The neutron separation energies of the xenon isotope chain predicted by the pseudo DRHBc and RCHB mass tables as functions of mass number $A$. (c) The abundance ratio of adjacent even-even nuclei along the xenon isotope chain predicted by the pseudo DRHBc and RCHB mass tables as functions of mass number $A$, under astrophysical conditions with neutron number density $n_n=$ $10^{24.5} \mathrm{~cm}^{-3}$ and temperature $T \gtrsim 1.5 \mathrm{~K}$. (d) Relative abundance distribution of even-even nuclei along the xenon isotope chain predicted by the pseudo DRHBc and RCHB mass tables as functions of mass number $A$, before the neutrons freeze-out.}
    \label{Xe}
\end{figure}

As is evident from Eq.~\eqref{saha}, the single neutron separation energy $S_n$ manifests in the exponent, rendering it the primary quantity that has a direct impact on the $r$-process simulations.
Taking the neutron flow with $n_n = 10^{24.5}$ cm$^{-3}$ and the xenon ($Z=54$) isotopic chain as examples, we analyze the deformation effect on the $S_n$ of $r$-process nuclei around $A=160$. 
The xenon isotopes show prolate shapes at $A=140-159$ and oblate shapes for $A=160-170$, as shown in Fig.~\ref{Xe}(a). 
Figure \ref{Xe}(b) shows the predicted neutron separation energies for the xenon isotopic chain based on the pseudo DRHBc and RCHB mass tables as functions of $A$.
After including the deformation effect, the neutron separation energies predicted by the pseudo DRHBc mass table decrease more rapidly with $A$.
Note that in the present $r$-process simulations, the neutron flow with neutron number density $n_n = 10^{24.5}$ cm$^{-3}$ contributes the most to the abundance near $A=160$.
According to Eq.~\eqref{saha}, as shown in Fig.~\ref{Xe}(c), the abundance ratios of adjacent even-even nuclei predicted by the pseudo DRHBc mass table decrease more rapidly with $A$. 
Additionally, as the mass number $A$ increases, the abundance ratios of adjacent even-even nuclei decrease from several orders of magnitude above 1 to below 1. 
This indicates that as the neutron number increases, the abundance of even-even nuclei gradually increases until reaching a maximum and then decreases. 
Figure \ref{Xe}(d) shows the relative abundance distribution of even-even nuclei on the xenon isotopic chain. 
It is found that, compared to the results of the RCHB mass table, the maximum of the abundance distribution, which corresponds to the $r$-process path nuclei, based on the pseudo DRHBc mass table is located at smaller mass numbers. 
This is the reason that the $r$-process abundances based on pseudo DRHBc mass table are higher for $A=148-165$ compared to those based on the RCHB mass table, as shown in Fig.~\ref{r_abundance}.

\begin{figure}[htbp]
	\centering
	\resizebox{\linewidth}{!}{\includegraphics{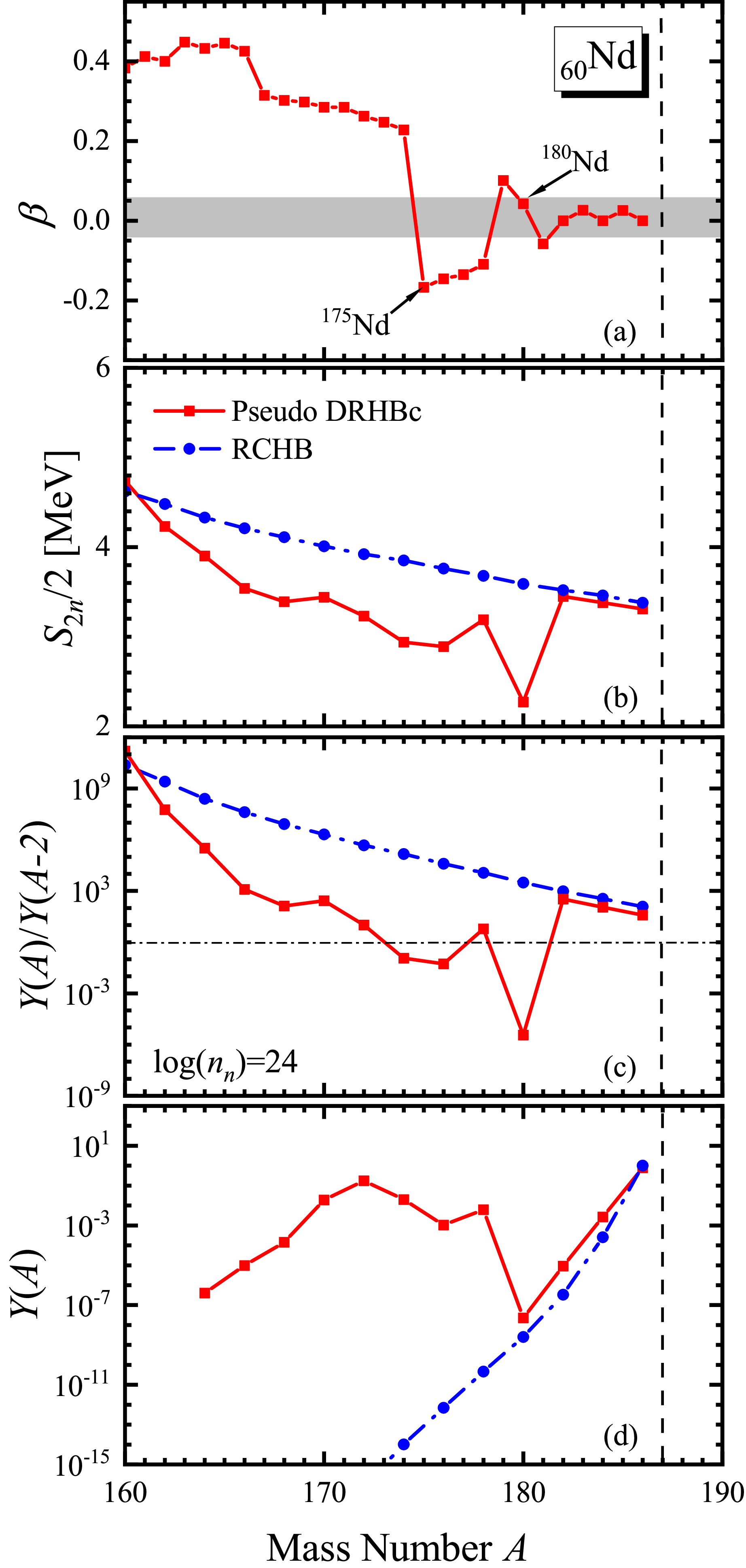}}
	\caption{(a) The nuclear quadrupole deformations $\beta_2$ of neodymium isotopes predicted by the DRHBc
    theory as a function of mass number $A$. (b) The neutron separation energies of the neodymium isotope chain predicted by the pseudo DRHBc and RCHB mass tables as functions of mass number $A$. (c) The abundance ratio of adjacent even-even nuclei along the neodymium isotope chain predicted by the pseudo DRHBc and RCHB mass tables as functions of mass number $A$, under astrophysical conditions with neutron number density $n_n=$ $10^{24} \mathrm{~cm}^{-3}$ and temperature $T = 1.5 \mathrm{~K}$. (d) Relative abundance distribution of even-even nuclei along the neodymium isotope chain predicted by the pseudo DRHBc and RCHB mass tables as functions of mass number $A$.}
    \label{Ne}
\end{figure}

As for the $r$-process nuclei around $A=175$, the neutron flow with neutron number density $n_n=10^{24}$ cm$^{-3}$ and the neodymium ($Z=60$) isotopic chain are taken to analyze the impact of deformation effect. 
Figures \ref{Ne}(a) and (b) represent the predicted quadrupole deformations $\beta_2$ and neutron separation energies for the neodymium isotopic chain, respectively. 
It is found that the pseudo DRHBc mass table with rotational corrections predicts a sudden decrease in neutron separation energy at $A=180$. 
This sudden change arises because the DRHBc calculation neglects rotational corrections for nuclei with $|\beta_2| < 0.05$ since the cranking approximation used is not suitable for near-spherical nuclei~\cite{Zhang2022ADNDT,Guo2024ADNDT}, as has been discussed in Section \ref{Stheory}. 
As shown in Fig.~\ref{Ne}(a), the quadrupole deformations $\beta_2$ from the DRHBc calculations for neodymium isotopes drop below 0.05 at $ A = 180 $, causing the sudden reduction in neutron separation energy for this isotope. 
Consequently, this sharp decline reduces the relative abundance distribution at $A = 180$. 
Following the neutron freeze-out stage, this phenomenon influences the $r$-process abundance pattern within the $A = 170-178$ region, primarily as a result of the $\beta$-decay processes that are accompanied by neutron emission.
In addition, it is found that the neutron separation energies predicted by the pseudo DRHBc mass table show a non-monotonic behavior as $A$ increases and represent a bump around  $A=175-182$.
This leads to abundance ratios of adjacent even-even isotopes, as shown in Fig.~\ref{Ne}(c), being less than 1 around $A = 172-176 $ and larger than 1 for $A>176$, thus forming a minimum near $A=176$. 
Correspondingly, the relative abundances predicted by the pseudo DRHBc mass table exhibit a trough around $A=176$, as shown in Fig.~\ref{Ne}(d). 
It is noted in Fig. \ref{Xe}(a) that Xe isotopes also show $|\beta_2|$ crossing $0.05$ near $A=180$, which leads to a trough in $S_{2n}/2$ and $Y(A)/Y(A-2)$, but since their $Y(A)/Y(A-2)$ values are significantly smaller than $1$ and $A=180$ reaches the neutron drip line for Xe, the abundances of such extremely neutron-rich isotopes are very low, and therefore, the trough in abundance for Xe isotopes does not show up.
Similar trough has been noted in Ref.~\cite{Martin2016PRL}, where the abundance based on SLy4 mass model shows a trough around $A=180$, which does not appear in the results based on SkM* and UNEDF0 mass models. 
This trough was also attributed to the nonmonotonic behavior of $S_{2n}$, which was traced back to rapid shape transitions from prolate to oblate.
As shown in Fig.~\ref{Ne} (a), the quadrupole deformation $\beta_2$ for the neodymium isotopes shows a rapid shape transitions from prolate to oblate at $A=175$, corresponding to the predicted non-monotonic neutron separation energy changes.

The abundance trough around $A=170$ from classical $r$-process simulation with the pseudo DRHBc mass table is significant.
As it is very different from the observed solar $r$-process abundances, here we provide three more remarks: 
\begin{enumerate}
  \item[(i)] First, from the view of nuclear structure, the rapid shape transitions have led to the abundance trough around $A=170$.
  Rapid shape transitions occur due to the shape-coexistence of prolate and oblate configurations for nuclei in this region.
  Such shape coexistence might correspond to the triaxial deformation \cite{Zhang2023PRC_TRHBc,Zhang2025arXiv}. 
  Incorporating the triaxiality is expected to smooth the evolution of nuclear shapes, leading to consistently smoother evolutions of $S_{2n}$ and $Y(A)$ and resolving the abundance trough. 
  In addition, the shape-mixed beyond-mean-field effects \cite{Sun2022CPC} could also potentially improve the abundance description at this trough.
  
  \item[(ii)] Second, from the view of $r$-process simulation, the present simulations are performed using the classical $r$-process model with some approximations.
  The sudden neutron freeze-out approximation could lead to an unrealistic final abundance distribution. 
  In the dynamical $r$-process studies, the competition between neutron capture and $\beta$-decay during the gradual neutron freeze-out stage can redistribute the abundance distribution and smooth local troughs to some extent.
  In this sense, this abundance trough could be partially eliminated in the full-network dynamical $r$-process simulations.
  
  \item[(iii)] 
  Third, from the view of methodology, it has been concluded above that the $r$-process abundances are not very sensitive to the details of pairing gaps. 
  Therefore, it is reasonably inferred that this trough might still exist in the abundance based on the full DRHBc mass table in the future, due to the shape transitions. 
  One can expect the examination for this trough after the release of the full DRHBc mass table.
  
\end{enumerate}

\section{Summary} \label{Ssum}

In summary, based on the available even-$Z$ part of the DRHBc mass table with the density functional PC-PK1 and the density-dependent zero-range pairing force, the masses of odd-$Z$ nuclei are estimated by approximating the odd-even mass differences with microscopic average pairing gaps.
A pseudo DRHBc mass table is constructed, and applied to $r$-process simulation.

The estimation is realized by expressing the mass of an odd nucleus as a function of its neighboring even ones' masses and odd-even mass differences, where the latter are approximated by the microscopic average pairing gaps. 
This treatment is examined by taking all even-odd nuclei with $8 \leqslant Z \leqslant 120$ as examples and estimating their binding energies based on the DRHBc results of even-even nuclei. 
It is found that the estimated binding energies of even-odd nuclei can effectively reproduce the self-consistent DRHBc calculation results. 
The comparisons with the results based on the empirical pairing gaps $12A^{-1/2}$ and $34A^{-3/4}$ show that the global behavior of the binding energy is not very sensitive to the details in the evolution of pairing gaps.
Then we estimate the binding energies of odd-$Z$ nuclei based on the DRHBc results of even-$Z$ nuclei, and combine them to construct a pseudo DRHBc mass table for all the nuclei with $8\leqslant Z \leqslant 120$. 
The rms deviation of binding energy from available data $\sigma=2.55$ MeV (without rotational correction) and 1.47 MeV (with rotational correction). 
The one/two-neutron/proton separation energies as well as their accuracies are also calculated and found to be close to the real DRHBc results from self-consistent calculations. 

The $r$-process simulation is performed based on the obtained pseudo DRHBc mass table. 
The pairing gaps chosen are sufficient to support the estimation of masses for nuclei that have not been explicitly calculated.
The impact of nuclear deformation effect is analyzed via comparing with results from the RCHB mass table.
The simulations from both mass tables predict abundance peaks near $A = 130,195$ and an abundance trough around $A = 170$. 
Compared to the RCHB mass table, the pseudo DRHBc mass table predicts higher abundances in $A=148-165$ and lower abundances in $A=170-180$. 
These differences can be explained by the $r$-process path nuclei predicted by the pseudo DRHBc mass tables; namely, around $A=160$, they are closer to the stability line, while around $A=175$ there is a gap.
It is found that the differences in the $r$-process path nuclei around $A=160$ are caused by the faster decrease of neutron separation energies predicted by the pseudo DRHBc mass table along mass number $A$ on the $Z=54-58$ isotope chains. 
The absence of $r$-process path nuclei around $A=175$ is attributed to the non-monotonic behavior of neutron separation energy and also influenced by the treatment of rotational corrections in the present DRHBc calculations.
Detailed analysis indicates that both of these two differences originate from the deformation effects.

For perspective, full-network dynamical $r$-process simulations based on the upcoming full DRHBc mass table will be essential to clarify the abundance trough around $A=170$.
Incorporating the triaxial deformation and the beyond-mean-field effects in the DRHBc calculations for several key nuclei are also crucial, as more sufficient descriptions for nuclei with rapid shape transitions could be provided. 
Further improvements are expected based on these extensions. 
In addition, a more detailed comparison with other microscopic or semi-empirical mass models in terms of $r$-process abundance results is to be conducted in future investigations.

\section*{Acknowledgments}

Helpful discussions with members of the DRHBc Mass Table Collaboration are highly appreciated. 
This work was partly supported by the 
	National Natural Science Foundation of China under Grants No. 12405134, No. 12435006, No. 12141501, and No. 12475117, No. 11935003, 
	the State Key Laboratory of Nuclear Physics and Technology, 
	Peking University under Grants No. NPT2023KFY02, and  No. NPT2023ZX03, 
	the National Key Laboratory of Neutron Science and Technology under Grant No. NST202401016, 
	National Key R\&D Program of China 2024YFE0109803, 
	High-performance Computing Platform of Peking University, 
	the China Postdoctoral Science Foundation under Grant No. 2021M700256, 
	and the start-up Grant No. XRC-23103 of Fuzhou University.

%

\end{document}